\documentclass[sigconf,review]{acmart}
\usepackage{multirow}

\usepackage{amsmath,amssymb,amsfonts}
\usepackage{algorithmic}
\usepackage{graphicx}
\usepackage{textcomp}
\usepackage{xcolor}
\usepackage{amsthm}
\usepackage{url}
\usepackage{booktabs}
\usepackage{array}
\usepackage[linesnumbered]{algorithm2e}
\usepackage{subcaption} 
\usepackage{multirow}
\usepackage{adjustbox}
\usepackage{booktabs}

\newtheorem{Definition}{Definition}
\newcommand{\one}{({\em i}\/)\xspace}
\newcommand{\two}{({\em ii}\/)\xspace}
\newcommand{\three}{({\em iii}\/)\xspace}
\newcommand{\four}{({\em iv}\/)\xspace}
\newcommand{\five}{({\em v}\/)\xspace}
\AtBeginDocument{%
  }

\setcopyright{none}
\renewcommand\footnotetextcopyrightpermission[1]{} 
\begin{document}

\title{SLIM: a Scalable Light-weight Root Cause Analysis for Imbalanced Data in Microservice}


\author{Rui Ren}
\email{renrui2019@ict.ac.cn}
\affiliation{%
  \institution{DAMO Academy, Alibaba Group Hangzhou, China}
}

\author{Jingbang Yang}
\email{jingbang.yjb@taobao.com}
\affiliation{%
  \institution{DAMO Academy, Alibaba Group Hangzhou, China}
}
\author{Linxiao Yang}
\email{linxiao.ylx@alibaba-inc.com}
\affiliation{%
  \institution{DAMO Academy, Alibaba Group Hangzhou, China}
}
\author{Xinyue Gu}
\email{guxinyue.gxy@alibaba-inc.com}
\affiliation{%
  \institution{DAMO Academy, Alibaba Group Hangzhou, China}
}
\author{Liang Sun}
\email{liang.sun@alibaba-inc.com}
\affiliation{%
  \institution{DAMO Academy, Alibaba Group Hangzhou, China}
}

\begin{abstract}The newly deployed service - one kind of change service, could lead to a new type of minority fault. Existing state-of-the-art methods for fault localization rarely consider the imbalanced fault classification in change service. This paper proposes a novel method that utilizes decision rule sets to deal with highly imbalanced data by optimizing the F1 score subject to cardinality constraints. The proposed method greedily generates the rule with maximal marginal gain and uses an efficient minorize-maximization (MM) approach to select rules iteratively, maximizing a non-monotone submodular lower bound. Compared with existing fault localization algorithms, our algorithm can adapt to the imbalanced fault scenario of change service, and provide interpretable fault causes which are easy to understand and verify. Our method can also be deployed in the online training setting, with only about 15\% training overhead compared to the current SOTA methods. Empirical studies showcase that our algorithm outperforms existing fault localization algorithms in both accuracy and model interpretability.
\end{abstract}

\maketitle

\section{Introduction}

Monolithic services have been progressively restructured into more refined modules, comprising of hundreds (or even thousands) of loosely-coupled microservices~\cite{seer,deathstar,suite,intro1}. Leading companies like Netflix, eBay and Alibaba have adopted this application model. Microservices offer several benefits that make them a powerful architecture, including simplification of application development, resource provisioning efficiency and flexibility. Despite these promising advantages, microservices introduce complex interactions among modular services, which can make on-demand resource provisioning challenge and potentially lead to performance degradation.

To enable engineers to resolve failure efficiently, fault localization is at the core of software maintenance for online service systems. With the advancement of monitoring and collecting tools, Metrics, Traces, and Logs have become the three fundamental elements of fault localization. Metrics refer to numeric data measurements taken at regular intervals of time, which help to understand the reasons behind the functioning of your application. Each trace records the process of a request being called through service instances and their operations~\cite{intro7,intro8}. Logs provide detailed information on the system's running status and user behavior. 

Though tremendous efforts have been devoted to software service maintenance and observability, in practice failures are inevitable due to the the increasing size and complexity of systems, which can result in significant economic losses and user dissatisfaction~\cite{intro1,intro2,intro3}. Analyzing the root causes of such performance issues is non-trivial and often error-prone, as hundreds of services may exhibit anomalies (e.g., network congestion and limited available cores) and propagate to dependent services. Moreover, large microservice systems are highly active and dynamic~\cite{intro1}, with numerous service changes.

It is important to note that the faults in a service often occur within the change of service, e.g. initial period of deploying new service or code change of services. This is due to changes in system architecture, service version (with backward or forward compatibility), and insufficient testing of the new service itself~\cite{ver1,ver2}. Google SRE book points out that about 70\% of failures are caused by changes in services~\cite{ver3}. Additionally, the lack of adequate fault data for newly deployed services hinders the learning process, making root cause analysis algorithms hard to localize such faults, even leading to a chain of incidents. However, existing fault localization algorithms mainly focus on utilizing new deep-learning and machine-learning models to fully learn the information in the multi-source data (log, metric and trace), while ignoring the highly dynamic running-time status with numerous services in change and the limited training data. Thus, they are unable to quickly learn and respond to new faults caused by service changes with a small amount of fault data~\cite{icse6,icse7}.

\textbf{Motivations.} Existing fault localization methods for imbalanced classification generally rely on re-sampling the training data~\cite{tracerca,dejavu}. In fact, simply over-sampling the minority class samples or down-sampling the majority class samples may cause model overfitting and on the other hand cannot generate extra insights from the data. We show that those SOTA algorithms cannot achieve significant performance improvement through re-sampling in our experiment part~\ref{imb_exp}. Besides, many deep learning-based fault localization methods~\cite{seer,sage,LLM,icse8} which claim to have good fault localization performance, generally require several hours for a single training session. Thus, they fail to handle services experiencing frequent failures. In addition, those algorithms only offer simple binary classification results with a lack of interpretability, making it difficult for engineers to understand, diagnose, and further prevent faults in the next steps. 
Therefore, how to build an interpretable fault localization model that could quickly and accurately respond to newly deployed service fault patterns from a limited number of imbalanced failures is an appealing but challenging problem~\cite{icse1,icse2}.

We summarize three main technical challenges to build effective fault localization models for imbalanced datasets of newly deployed services as follows.

\begin{itemize}
    \item[1)] The first challenge arises from imbalanced data of service change (e.g., newly deployed service). We also note that simply applying re-sampling does not lead to performance improvement in our experiments. 
    \item[2)] The second challenge is the model interpretability for operating engineers. Good interpretability can help engineers to better understand the fault cause and possibly identify related risks. Unfortunately, most of the existing works on fault localization lacks interpretability. 
    \item[3)] The third challenge arises from the unbearable training overhead of existing models, primarily attributed to the dynamic runtime environment in the microservice scenario.
\end{itemize}


In this paper, we aim to design an interpretable classifier on highly imbalanced data via learning decision rule sets~\cite{CG-Dash} for fault localization. It can be expressed in disjunctive normal forms~(DNF, OR-of-ANDs), which enjoys good interpretability due to the logical clauses. An example of DNF models with two conditions is ``IF (cpu\_usage$>$80 AND file\_disk\_read$>$180) OR (file\_disk\_write$>$70 AND memory\_usage
$>$170) THEN $\hat{y}=1$''. It helps engineers to understand which key metrics are affected by the fault. As we only focus on building rules for the minority class (further called the positive class), thus it is an ideal interpretable model for imbalanced classification. Moreover, our model can achieve incremental (or online) training within minutes for every newly deployed services, which makes it deployable to a wide range of services with low cost. 

In this paper, we propose a fault localization algorithm called \textbf{SLIM}~(\textbf{S}calable and interpretable \textbf{L}ight-weight algorithm for \textbf{I}mba\-lanced data in \textbf{M}icroserverice) to address the aforementioned challenges. Here we summarize the main contributions as follows: 
\begin{itemize}
    
    \item[1)] To the best of our knowledge, \textbf{SLIM} is the first fault localization algorithm to address the issue of imbalanced fault data in service change (e.g., newly deployed services) from an algorithm-level within the microservices environment. 

    \item[2)] Our fault localization algorithm can generate the interpretable rule set that could assist engineers in understanding the root causes of failures. 

    \item[3)] Our \textbf{SLIM} is efficient and can be deployed easily with a low cost. Compared to other algorithms, our model's training time is only around 15\% of theirs in the most complex scenarios. 
    \item[4)] We apply our model's interpretable ruleset to two use cases, which replace human experts to build prior knowledge. The results show that our interpretable ruleset is comparable to expert knowledge and reduces 80\% of the time required. Besides, our ruleset knowledge base beats the precision of other interpretable methods' knowledge base.

    \item[5)] We have conducted extensive experiments on 359 failures from three systems including three open-source benchmark datasets~\cite{dejavu}. The results show SLIM's effectiveness, efficiency and interpretability. We also apply our algorithm in the real running-time environment of the largest cloud service system provider in China. We give the detailed case study in the section of experiment. We also provide the demo available on the github~\footnote{https://anonymous.4open.science/r/SLIM-5B7F/}.

\end{itemize}



\section{The SLIM Algorithm}

\subsection{The Pipeline of SLIM}
SLIM is an interpretable and scalable fault localization algorithm for microservices system. It identifies the (faulty) services that cause performance degradation in a microservice system, and provides explanations for operators to understand why. Figure~\ref{fig:pipeline} shows an overview of SLIM's pipeline, primarily involving 4 modules. Firstly, it processes logs and converts them into metrics as features. Then, the features are transformed into binary encoders. With these binary features as inputs, rules can be learned. Finally, the learned rules are used to vote out faults. We will now introduce these 4 modules in detail.

\subsubsection{Log Extraction Module}
\begin{figure}
    \centering
    \includegraphics[width=1\linewidth]{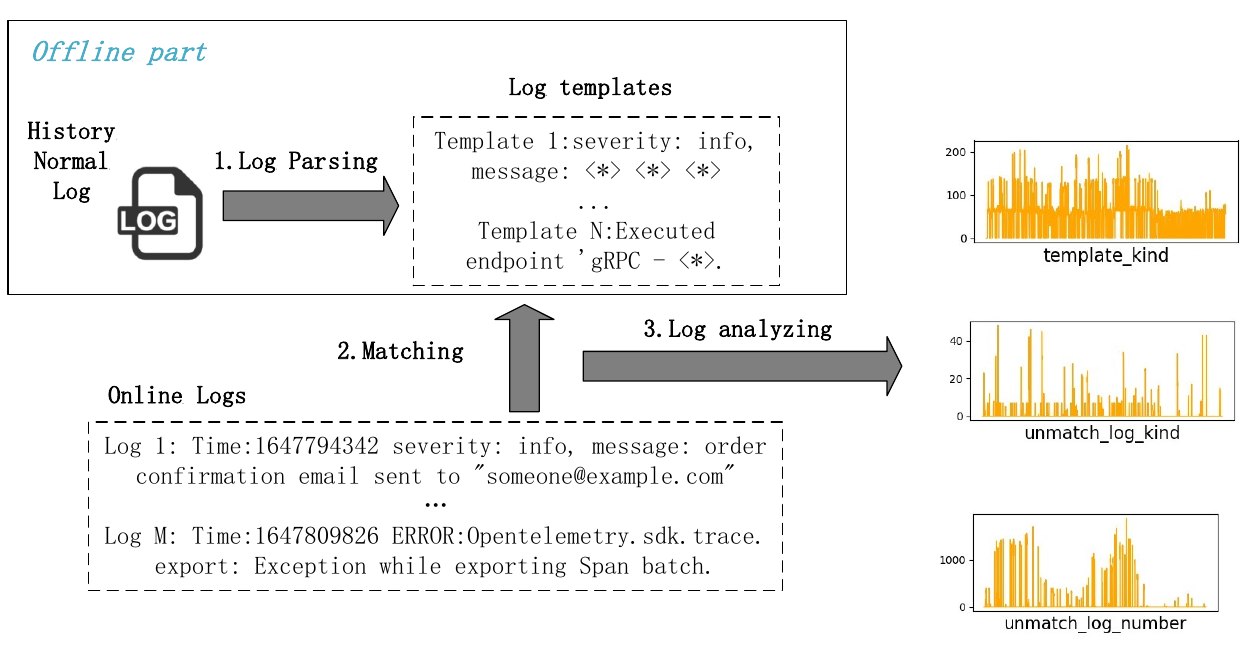}
    \caption{Log Extraction Module: Log Parsing, Matching and Analyzing.}
    \label{fig:logparse}
    \vspace{-15pt}
\end{figure}

This module extracts key log information and converts it to metric data. Operational log data is in unstructured format and not suitable for direct training. So we leverage log extraction methods~\cite{scwarn} to deal with it. As Figure~\ref{fig:logparse} shows, the procedure consists of log template parsing and analysis of template variation. We first parse the normal history log messages to construct a standard template base offline using drain~\cite{drain}. These normal log templates serve as a historical reference to identify whether any new template exists in online log messages. Then we parse the online log messages and construct streaming time-series log templates. We compare the streaming time-series log templates with the template base and record the unmatched log template, along with their quantity, types, and other features. We aggregate these features by time interval, aligning them with the metrics' sample interval.

\subsubsection{Feature Binarization Module}
This module generates binarized features from the metrics obtained in the feature extraction module. We employ a bucketing strategy using a sequence of thresholds to cut the numerical features into discrete, binarized features. For example, in Figure~\ref{fig:pipeline}, the network latency has a continuous distribution between 100-500ms. The data is discretized into multiple values, such as $Network\_Latency \leq 100$, $100< Network\_Latency < 200$ and $Network\_Latency \geq 200$. For categorical features, we use one-hot encoding to generate binarized features. We apply this discretization process uniformly across the data distribution. These discrete values can be combined to form rules.

\subsubsection{Efficient Rule Learning Method}
\label{sec:efficient_rule}
This module digs out the rule set by our proposed novel classifier. The algorithm first identifies rule sets for each fault type. Namely, for each fault type we train a sub-model and obtain a rule set. Then, those which hit the rules in the rule set are classified as corresponding fault samples. The detailed rule learning method will be discussed in Section~\ref{sec:detail}.

\subsubsection{Fault Localization}\label{sec:faultloc}
This module provides the final results, including the localization of fault types and the localization of fault services. We design fault ranking methods by adopting a voting mechanism that takes into account both the hit counts of each rule set and the rule's confidence.

\subsection{Detailed Procedure of Rule Set Learning}\label{sec:detail}
In the section, we introduce our rule learning methods, including Rule Set Selection in Section~\ref{rulesel} and Efficient Rule Generation in Section~\ref{rulegen}. Before diving into the details, we first predefine some key notations.

\subsubsection{Notations and Preliminaries}
Given a dataset $\mathcal{X}=\{(\boldsymbol{x}_i, y_i)\}_{i=1}^n$, $\boldsymbol{x}_i\in\{0,1\}^{d}$ is 
the binary feature vector obtained from feature binarization and $y_i\in\{0,1\}$ is the true label indicating the belongingness of a given fault type. Here, $d$ is the size of the feature index set $\Gamma$. A sample $(\boldsymbol{x}_i, y_i)$ is positive if
$y_i=1 (abnormal)$ and negative if $y_i=0 (normal)$. We call the $i$-th sample is covered by feature $j$ if $x_{i,j}=1$. For example, given  $j\in d$, the $j$-th feature $Network\_delay > 200ms$, $x_{i,j}=1$ means for the $i$-th sample, $Network\_delay > 200ms$ and $x_{i,j}=0$ means $Network\_delay \leq 200ms$. Denote $h_i\in\{0, 1\}$ as the prediction of the
$i$-th sample.
For a certain fault type, we aim to predict $y$ from the dataset $\mathcal{X}$ using an interpretable rule set.

\textbf{Rule and rule set.} Our classifier is a rule set that consists of rules. We now give the definitions of these two ingredients and the related notations.
\begin{Definition}[\textbf{Rule}]\label{def:rule}
A rule $r$ is a set of feature indices, i.e. $r$ is a subset of ~$\Gamma$.
A sample $(\boldsymbol{x}_i, y_i)$ is covered by a rule $r$ if and only if $r \subseteq \{j\in \Gamma| x_{i,j}=1\}$.
\end{Definition}
\begin{Definition}[\textbf{Rule Set}]\label{def:rule set}
A rule set $s$ consists of multiple rules, and serves as a classifier,
which classifies a sample as positive if the sample is covered by at 
least one rule in $s$, and as negative if there is no rule
in $s$ that covers it. Therefore, the predicted label $h_i$ can be calculated via $\bigvee_{r\in s}(\bigwedge_{j\in r}x_{i,j})$.
\end{Definition}

Let $\mathcal{X}_j$, $\mathcal{X}_r$ and $\mathcal{X}_s$ denote the set of samples covered by the $j$-th feature, the rule $r$ and the rule set $s$,
respectively. In the following we use these different subscripts to distinguish sets of samples covered by different features/rules/sets, i.e. $\mathcal{X}_j=\{i|\boldsymbol{x}_{i,j}=1\}$, $\mathcal{X}_r=\{i|(\bigwedge_{j\in r}\boldsymbol{x}_{i,j})=1\}$, $\mathcal{X}_s=\{i|\bigvee_{r\in s}(\bigwedge_{j\in r}\boldsymbol{x}_{i,j})=1\}$.
According to the relationships among the features, rules and rule sets, we get $\mathcal{X}_r=\bigcap_{j\in r}\mathcal{X}_j$
and $\mathcal{X}_s=\bigcup_{r\in s}\mathcal{X}_r$.
We define a set operator $^{+}$ as a positive sample filter, meaning that $\mathcal{X'}^+$ returns a set containing all positive samples in the arbitrary given set $\mathcal{X'}$.

\textbf{Problem Formulation.} The common evaluation metrics, accuracy and error rate, are no longer applicable in imbalanced classification as they are prone to be dominated by the majority class~\cite{Johnson2019:survey}. To address this issue, we choose \textbf{F1 score} instead, which combines both precision and recall and cares about the performance of the minor positive samples:
\begin{align}
    F1(s)=\frac{2\sum_i y_i h_i}{\sum_i h_i+\sum_i y_i}. 
\end{align}
Since $\sum_i y_i h_i$ is the number of correctly classified positive samples, $\sum_i h_i$ is the number of positively predicted samples, $\sum_i h_i$, $\sum_i  y_i h_i$ and $\sum_i y_i$ 
can be rewritten as $|\mathcal{X}_s|$, $|\mathcal{X}_s^{+}|$ and $|\mathcal{X}^+|$, respectively. Formally, we formulate the F1 score as follows: 
\begin{align}
    F1(s) = \frac{2 \vert \mathcal{X}_s^+\vert}{\vert \mathcal{X}_s\vert + \vert \mathcal{X}^{+} \vert} = \frac{2 \vert \cup_{r \in s} \mathcal{X}^+_r\vert}{\vert \cup_{r \in s} \mathcal{X}_r\vert + | \mathcal{X}^+ |}. 
    \label{eqn:f1_defination}
\end{align}
To maximize F1 score, we have the following optimization problem: 
\begin{align}
\begin{split}
\label{eqn:objective}
        \max_{s\subseteq\Omega}& \ \frac{2 \vert \cup_{r \in s} \mathcal{X}^+_r\vert}{\vert \cup_{r \in s} \mathcal{X}_r\vert + | \mathcal{X}^+ |}, \\
    \mbox{s.t. } & \ |s| \le K, \ \Omega=\{r||r|\le l, r\subseteq\Gamma\}. 
\end{split}
\end{align}
where we set a predefined parameter $l$ to be the maximum feature length of a rule, and $K$ to be the maximum number of rules in a rule set. These constraints are to ensure the interpretablity of rules. By taking the logarithm, we can rewrite the objective in the following form:
\begin{align}
    \max_{s\subseteq \Omega}\ \log (\vert \cup_{r \in s} \mathcal{X}_r^+\vert)-\log (\vert \cup_{r \in s} \mathcal{X}_r\vert + | \mathcal{X}^+| ).
\label{eqn:logobj}
\end{align}

\subsubsection{Rule Set Selection}\label{rulesel}
We now present the details of our efficient rule selection method. Let us rewrite the two logarithm components of \eqref{eqn:logobj} as $G(s)\triangleq\log (\vert \cup _{r \in s} \mathcal{X}_r^+\vert)$ and $C(s)\triangleq\log (\vert \cup_{r \in s} \mathcal{X}_r\vert + |\mathcal{X}^{+}|)$. 
As logarithm function is non-decreasing and concave,
both $G(s)$ and $C(s)$ are non-negative monotone submodular functions~\cite{bach2013learning}. Consequently, the objective function can be viewed as a difference between two submodular functions. 
Our proposed method, which is referred as SLIM, is based on the method \textit{DistortedGreedy} \cite{harshaw2019submodular}, for maximizing
the difference between a non-negative monotone submodular function 
and a modular function. We will show that
by introducing the notation curvature, \textit{DistortedGreedy} is
applicable to our problem. We first define the curvature $\gamma$
of $C(s)$
\begin{align}
\gamma \triangleq 1-\min_{r\in\Omega}\frac{C(r|\Omega\setminus\{r\})}{C(r|\emptyset)}
\end{align}
to measure the closeness of $C(s)$ to a modular function, and $\gamma$ is unknown a priori.
The complete procedure for rule set selection is summarized in Algorithm \ref{alg:selectrules}.
Given the training dataset, the maximal number of rules and the limitation on the length of rules,
and by initializing the rule set as an empty set, we iteratively add a rule $r^{*}$ maximize a distorted marginal gain of $r$ with a parameter $\alpha$, i.e.,
\begin{align}
\label{eq:gc-1}
\max_{r}\quad \alpha G(r|s)-C(r|s), 
\end{align}
where $G(r|s)\triangleq G(s\cup \{r\})-G(s)$ and $C(r|s)\triangleq C(s\cup \{r\})-C(s)$ denote the marginal gains of $G$ and $C$ when adding $r$ to $s$, which is stated in line 6 of Algorithm \ref{alg:selectrules}. 


\RestyleAlgo{ruled}
\SetKwInOut{Input}{Input}
\SetKwInOut{Output}{Output}
\SetKw{Define}{Define}
\SetKw{Return}{Return}
\SetKw{Initialize}{Initialize}

\begin{algorithm}[ht]
\fontsize{10}{12}\selectfont
\caption{Rule Set for Imbalanced Data Set}
\label{alg:selectrules}
\textbf{Input :}Training data $\{(\boldsymbol{x}_i, y_i)\}_{i=1}^n$, cardinality $K$, curvature $\gamma$,
maximal size of a rule $l$.\\
\textbf{Output :}Rule set $s$.\\
Let $s_0 \leftarrow \emptyset$\;
\For{$i = 0,1,...,K-1$}{
    $\alpha_i \leftarrow (1-\frac{\gamma}{K})^{K-(i+1)}$\;
    $r^{*} \leftarrow \arg\max_r\; \alpha_i G(r|s)-C(r|s)$\;
    \If{$\alpha_i G(r^*|s)-C(r^*|s)>0$}{
        $s_{i+1} \leftarrow s_i \cup \{r^*\}$\;
    }
}
\Return{$ s_{K-1}$} 
\end{algorithm}


Similar to \textit{DistortedGreedy}, we adaptively update the trade-off
between $G(r|s)$ and $C(r|s)$ using $\alpha$ in line 5 of Algorithm \ref{alg:selectrules}. 
In early stages, a small value of $\alpha$ is adopted to select rules with higher \textit{precision}. The value of $\alpha$ is gradually increased to improve the \textit{recall} of the rule set.
In other words, SLIM tends to select the rules with higher \textit{precision} in the early iteration steps and focuses on the rules with higher \textit{recall} later.
The rationale behind the $\alpha$ updating strategy is that when first focusing on the rules with
high \textit{precision} and low \textit{recall}, SLIM can achieve higher precision and later improve the recall by including more rules. However, if the rules with high \textit{recall} but low \textit{precision} are given more priority in early stages, it is difficult to eliminate the effect of the false positive samples.

\begin{figure}
    \centering
    \includegraphics[width=0.9\linewidth]{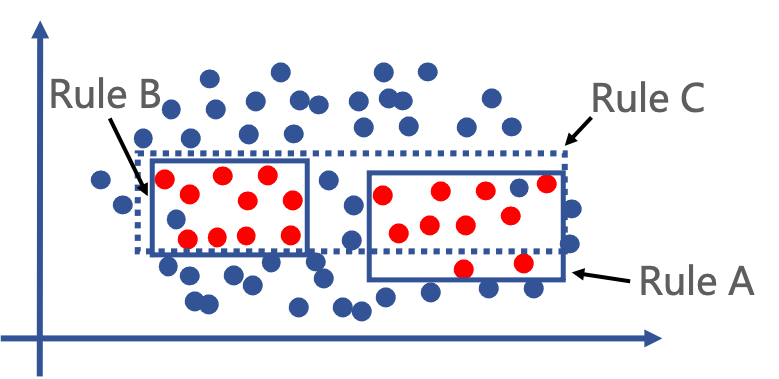}
    \caption{Example of the proposed rule selection strategy.}
    \label{fig:toy_example}
\end{figure}

To better illustrate this, we show a toy example in Figure~\ref{fig:toy_example}.
Given a dataset with 20 positive and 100 negative samples, and our goal is to select 2 rules from 3 candidate rules, namely rules A, B and C, where rule A covers 10 positive and 1 negative samples, rule B covers the rest 10 positive samples which are not covered by rule A and 1 additional negative sample, and rule C covers 18 positive samples and 5 negative samples.
We first discuss the scenario that we replace the $\alpha$ update
strategy in line 3 of Algorithm \ref{alg:selectrules} with $\alpha_i=1$.
At the first iteration for rule A, $\vert \mathcal{X}^+\vert=20$, $\vert \cup_{r \in s} \mathcal{X}_r\vert=11$, $\vert \cup_{r \in s} \mathcal{X}_r^+\vert=10$, then the marginal gain of rule A is $\log(10/(20+11))\approx \log(0.31)$.
Similarly, the marginal gains of rule B and rule C are given as $\log(10/(20+11))\approx \log(0.31)$, $\log(18/(20+18+5))\approx \log(0.42)$, respectively.
In this scenario, SLIM will select rule C in the first iteration and rule B(or A) in the second iteration. Finally, SLIM constructs 
a rule set which covers 20 positive samples and 6 negative samples.
However, with the proposed $\alpha$ update
strategy in line 3 of Algorithm \ref{alg:selectrules},
at the first iteration (corresponding to $K=2$, $i=0$, $\gamma=1$, and $\alpha=0.5$),
the marginal gains of rule A, B and C are given as $0.5 \times \log(10)-\log(31)(\approx \log(0.102))$, $0.5 \times \log(10)-\log(31)(\approx \log(0.102))$ and $0.5 \times \log(18)-\log(43)(\approx \log(0.099))$, respectively.
Then SLIM will return a better rule set that consists of rule A and rule B, which covers only $2$ negative samples.
We also give the theoretical guarantee for the proposed method at appendix~\ref{theorem:guarantee_1}


\begin{figure}
    \centering
    \includegraphics[width=0.9\linewidth]{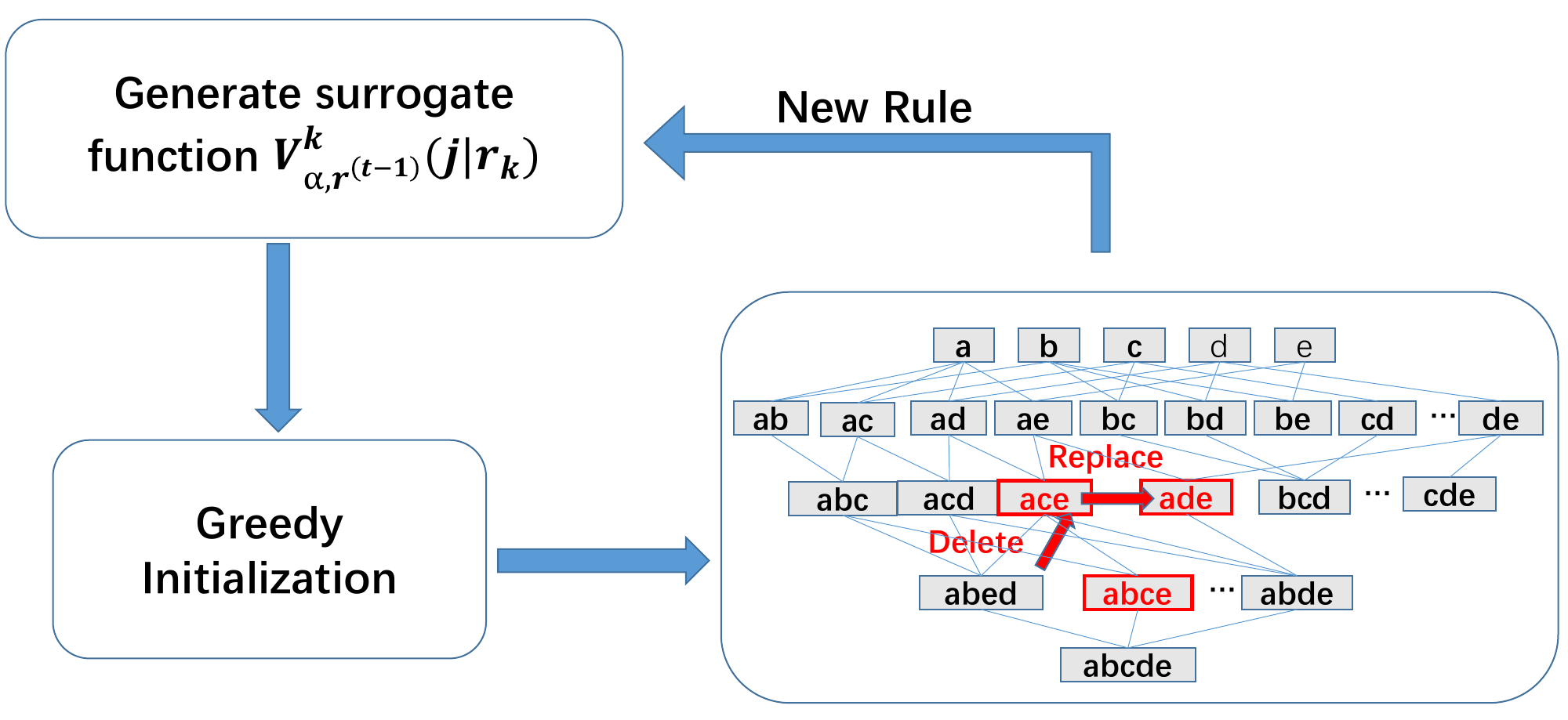}
    \caption{Framework of Rule Generation.}
    \label{fig:alg1}
    \vspace{-15pt}
\end{figure}

\subsubsection{Efficient Rule Generation}\label{rulegen}
Algorithm \ref{alg:selectrules} involves 
finding a rule $r$ to maximize a distorted marginal gain of $r$, i.e. solving problem (\ref{eq:gc-1}). 
As the number of possible rules is exponential with number of features, solving problem (\ref{eq:gc-1}) is NP-hard.
To address this issue, we propose an efficient rule generation method, which solves (\ref{eq:gc-1}) approximately.
Because $s$ is independent of $r$, (\ref{eq:gc-1}) can be reduced to
\begin{align}
\max_{|r|\le l} \quad\alpha \log(|\mathcal{X}_r^{+}\cup \mathcal{X}_s^+|)-
\log(|\mathcal{X}_r\cup \mathcal{X}_s|+|\mathcal{X}^+|).\label{eqn:obj-subproblem}
\end{align}
Notice that a rule $r$ is a set of feature indices, so finding an optimal rule is equivalent to finding a set of features. The expression in \eqref{eqn:obj-subproblem} can be further rewritten as:
\begin{align}
    W(r)\triangleq & \alpha \log(f(r)) - \log(g(r))
\label{eqn-W}
\end{align}
where $f(r)\triangleq|(\cap_{j\in r}\mathcal{X}_j)^{+}\cup \mathcal{X}_s|$ and $g(r)\triangleq|(\cap_{j\in r}\mathcal{X}_j)\cup \mathcal{X}_s|+|\mathcal{X}^+|$.

Directly maximizing $W(r)$ is difficult.
Although $f(r)$
is a supermodular function, the presence of logarithm function 
makes the property of $\log(f(r))$
non-trivial. In our algorithm, $W(r)$ is maximized using MM algorithm~\cite{hunter2004tutorial},
which iteratively increases the value of the objective function 
by maximizing a tight lower bound.
We propose a proper lower bound of $W(r)$ by finding a lower bound of $\log(f(r))$
and an upper bound 
of $\log(g(r))$ separately. 

Motivated by the modular upper bounds
of submodular functions presented in \cite{nemhauser1978analysis,jegelka2011submodularity},
two proper lower bounds of $f(r)$ for all $r \subseteq\ \Gamma$ are given as

\begin{align}\label{eqn:lb4supmod1}
L^{1}_{f, r^{(t)}}(r)\triangleq
&f(r^{(t)})-\sum \limits_{j\in Q_1} f(j|r^{(t)} \setminus \{j\})+\sum \limits_{j \in Q_2} f(j|\emptyset)
\le f(r)\nonumber\\
L^{2}_{f, r^{(t)}}(r)\triangleq
&f(r^{(t)})-\sum \limits_{j\in Q_1} f(j|\Gamma \setminus \{j\})+\sum \limits_{j \in Q_2} f(j|r^{(t)})
\le f(r)\nonumber
\end{align}
where $r^{(t)}$ denotes the current estimation of $r$, $Q_1=r^{(t)} \setminus r$ and $Q_2=r \setminus r^{(t)}$. These
two inequalities hold for all possible $r^{(t)}$, and 
the equality is achieved when $r=r^{(t)}$. 
We find an upper bound of $\log(g(r))$ by utilizing the concavity of logarithm functions.
As $\log(x) \le \log(x_0) + \frac{1}{x_0} (x-x_0)$,
then a tight upper bound of $\log(g(r))$ is readily given as, which holds for any $r^{(t)}$,

\begin{align}
    \log(g(r))\le \log(g(r^{(t)}))+\frac{1}{g(r^{(t)})}(g(r)-g(r^{(t)})). 
\end{align}
Combining the bounds obtained above, we derive two tight lower bounds
of $W(r)$ for $|r|\ge 1$ as follows,
\begin{align}
W(r)\ge& \alpha \log(L_{f,r^{(t)}}^{1}(r))-\frac{g(r)}{g(r^{(t)})}=V^{1}_{\alpha,r^{(t)}}(r),\\
W(r)\ge& \alpha \log(L_{f,r^{(t)}}^{2}(r))-\frac{g(r)}{g(r^{(t)})}=V^{2}_{\alpha,r^{(t)}}(r).
\end{align}

The problem of maximizing $W(r)$ is translated 
to maximizing $V^{1}_{\alpha}(r|r^{(t)})$ and $V^{2}_{\alpha}(r|r^{(t)})$. To step further, we give the detailed procedure of translation at the appendix~\ref{app:bound}.

Maximizing a non-monotone submodular function subject to cardinality constraints has been extensively studied in the literature.
Specifically, SLIM maximizes $V^{1}_{\alpha}(r|r^{(t)})$ and $V^{2}_{\alpha}(r|r^{(t)})$ by using a simple local search method.
As shown in \cite{lee2010maximizing}, by identifying the cardinality constraint as a matroid constraint,
the local search method can provide at least $1/4$-approximation to the optimum.

\begin{lemma}\label{lemma:submoudularity-V}
Both $V^{1}_{\alpha}(r|r^{(t)})$ and $V^{2}_{\alpha}(r|r^{(t)})$ are
non-monotone submodular functions. 
\end{lemma}

Maximizing a non-monotone submodular function subject to cardinality constraints has been extensively studied in the literature.
Specifically, SLIM maximizes $V^{1}_{\alpha}(r|r^{(t)})$ and $V^{2}_{\alpha}(r|r^{(t)})$ by using a simple local search method.
As shown in \cite{lee2010maximizing}, by identifying the cardinality constraint as a matroid constraint,
the local search method can provide at least $1/4$-approximation to the optimum.

Fig.\ref{fig:alg1} shows the overall framework of our rule generation method. At the $t$th iteration, we first generate surrogate functions of $W(r)$, i.e. $V^{1}_{\alpha,r^{(t)}}(r)$ and $V^{2}_{\alpha,r^{(t)}}(r)$, according to current estimation $r^{(t)}$. Then we maximize $V^{1}_{\alpha,r^{(t)}}(r)$ and $V^{2}_{\alpha,r^{(t)}}(r)$ utilizing local search technique and arrive at a new estimation $r^{(t+1)}$.
Our method only involves the set operation and is hence a computational efficient method. 
The computational complexity of our method can be further improved by permitting early stopping, 
i.e, terminating the local search if no significant improvement is achieved by replacing features.

\subsection{Detailed procedure of Fault Localization}
In this section, we introduce the detailed procedure of fault localization. For the localization of fault type, let the $R^{*}$ denote the selected fault type result, $R$ the collection of all fault types' rule sets, $x_{i}$ the $i$-th sample in the current time period,  $r_{j}$ the $j$-th fault type and $X$ all samples. $H_{r_j}(x_i)$ is an indicator function that takes the value 1 if $x_i$ is hit by the rule set $r_j$ and 0 otherwise. If the $i$-th sample is hit by multiple rules in the $j$-th ruleset, we select the rule with the highest precision $Precision^{max}_{r_j}$ at the training set to hit the sample. So we calculate the probability of the $i$-th sample belonging to the $j$-th fault type as Equation \eqref{eq:probx} writes. Then in Equation \eqref{eq:rmax} we sum up the sample-level probability for each fault-type rule set $j$ and pick out the fault-type with the highest value as the root cause: 
\begin{align}
&P(r_{j}|x_{i})=Precision^{max}_{r_j}*H_{r_j}(x_i),\label{eq:probx}\\
&R^{*}=\underset{r_{j} \in R}{\arg\max}\sum_{x_{i} \in X}{P(r_{j}|x_{i})}. \label{eq:rmax}
\end{align}


For the localization of service, let the $S^{*}$ denote the selected service result and $X_{k}$ all samples in the $k$-th service. $x_m \in X_{k}$ is the $m$-th sample in the $k$-th service and $H_{r_j}(x_m)$ means whether $x_m$ is hit by ruleset $r_j$. We first group the samples by service. As Equation \eqref{eq:probx2} demonstrates, similar to fault type localization, the probability for each sample is the highest precision's rule in $r_j$ when these rules in $r_j$ hits the sample and otherwise 0. Then, we individually sum up the product of all samples hit by the rules in each service multiplied by the corresponding rule precision score according to Equation \eqref{eq:rmax2}. We sort the probability results of every service and choose the service with the highest value as the root cause.
\begin{align}  
&P(r_{j}|X_{k})=\sum_{x_m \in X_{k}}{Precision^{max}_{r_j}*H_{r_j}(x_m)}\label{eq:probx2},\\
&S^{*}=\underset{X_k \in X}{\arg\max}\sum_{r_{j} \in R}{P(r_{j}|X_{k})}.\label{eq:rmax2}
\end{align}

The fault localization module could help us to confirm the failure service and fault type of the failure. We give a detailed evaluation of the effect of our model in the experiment part.

\begin{figure}
    \centering
    \includegraphics[width=1\linewidth]{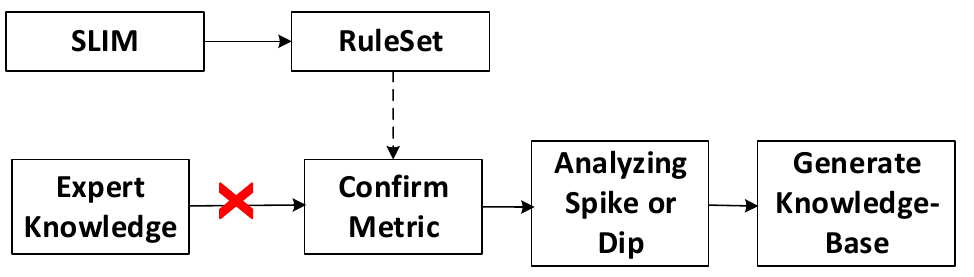}
    \caption{The overview of Knowledge Base generation}
    \label{fig:knowledgebase}
    \vspace{-20pt}
\end{figure}

\section{The Application of interpretable ruleset}
Our interpretable ruleset could assist engineers to confirm the root cause and find out the most relevant metric for the anomaly. It makes the model could corporate with expert knowledge for troubleshooting and debugging system errors. 
On this basis, we go a step further to apply our model to generate the knowledge base for existing fault localization algorithms. We will introduce these applications in the following part.

\subsection{Overview of the Knowledge Base Generator}
Usually, an anomaly knowledge base is constructed manually to store expert experience and quickly localize history fault types. When an anomaly firstly occurs, experts analyze its key features and generate fault fingerprints for the Knowledge Base. The fingerprint will help engineers to fast solve the problem again in the future.

Therefore, expert knowledge plays an important role for many algorithms~\cite{microcbr,cloudrca}. These algorithms leverage expert knowledge to describe every fault type and construct the prior knowledge from metrics and trace data. This is called the knowledge base or case base. Although the knowledge base is necessary for real systems, it is also expensive for the company to recruit experts and build a knowledge base manually. Thus, our interpretable ruleset can help to construct the knowledge base and reduce costs.
Fig. \ref{fig:knowledgebase} shows an overview of the knowledge base generation. We try to replace the expert knowledge component in these algorithms with our rule set in order to build the knowledge base required for diagnosing faults in their models.


\begin{table}[t]
\caption{Benchmark dataset statistics.}
\begin{center}
\begin{tabular}{c c c c c}
\hline
\textbf{Dataset}&\textbf{$\#$ Features}&\textbf{$\#$ Samples}&\textbf{$\#$ Service}&\textbf{$\#$ faults classes}\\
\hline
$\mathcal{A}$&124&62249&17&18\\
$\mathcal{B}$&501&412714&29&20\\
$\mathcal{C}$&8&469988&43&2\\
$\mathcal{D}$&12000&7200&2173&5\\
\hline
\end{tabular}
\vspace{-15pt}
\label{tab:fault_type}
\end{center}
\end{table}

\subsection{CloudRCA}
CloudRCA~\cite{cloudrca} leverages RobustSTL to extract the abnormal metrics and identify important system metrics using the expert knowledge base. The selected metrics are learnt via a Knowledge-informed Hierarchical Bayesian Network (KHBN) to perform root cause analysis.

In our implementation, we replace the RobustSTL and expert knowledge with our interpreter to find out the most important metric sequence. We first pass labeled training data through our SLIM, where the model analyzes the fault cases to give the key ruleset. Then, we construct the feature matrix according to the ruleset including the key metric and log information. Finally, the KHBN completes the root cause analysis. We finish the experiment with the A dataset and evaluate the CloudRCA's performance.

\subsection{MicroCBR}
Similarly, we also integrate our ruleset into MicroCBR~\cite{microcbr}, which leverage the labeled anomaly case to construct the knowledge base and perform root cause analysis through case-based reasoning. Each case records a specific root cause and its solution, along with a set of anomalies detected from resource metrics, logs and other operating information. The precision of case-based reasoning is always dominated by these abnormal metrics.

We leverage our ruleset to automatically construct anomaly knowledge for its knowledge-base. First, the SLIM to learn the labeled data. 
Then, our ruleset will show the important metrics for the all fault type. MicroCBR aims to record the time-series metrics' fingerprint, which includes every key metric's spike or dip. We leverage our ruleset to assist the MicroCBR to select the key metrics and construct fingerprints.

\section{Experiments}\label{sec:exp}


In this section, we perform experiments on benchmark datasets to show the performance of SLIM in comparison with the state-of-the-art fault localization algorithms.


\vspace{-10pt}
\subsection{Experimental Setup}

\subsubsection{Datasets} We conduct experiments on three public datasets, which are denoted as dataset $\mathcal{A}$,
$\mathcal{B}$~\cite{datasetAB},
$\mathcal{C}$~\cite{datasetC}, and $\mathcal{D}$, respectively.

Dataset $\mathcal{A}$ and $\mathcal{B}$ are obtained from two different production service systems, 
which are both injected in 18 types of faults that can be summarized as: 
\one CPU exhaustion on containers, physical servers and middleware. \two packet loss, delay on service and physical node. \three database connection limit and close(just for dataset $\mathcal{A}$); \four low free memory at JVM/Tomcat (just for dataset $\mathcal{B}$); \five Disk I/O exhaustion (just for dataset $\mathcal{B}$)~\cite{dejavu}. The dataset $\mathcal{C}$ is generated by the train ticket booking microservice system~\cite{trainticket,tracerca},
where the fault classes include the network delay and CPU consuming.  The dataset $\mathcal{D}$ comes from the real-world system that is one of the biggest cloud services provider(we refer to it as \textbf{Company ALC} for brevity). It consist of 35 incidents occurred in our cloud platform. These incidents are collected and verified which services are the root cause by our SRE team. For every fault, we have about 12000 metrics, collected during 3600 seconds(half hour before and half hour after the anomaly was reported) from 2173 microservices.
We summerize the number of features, number of samples,
number of services, and number of faults classes
in Table~\ref{tab:fault_type}.

\begin{table*}[t]
\caption{Accuracy comparison of different root cause localization algorithms.}
\begin{center}
\begin{tabular}{c c c c c c c c c c}
\hline 
\textbf{Dataset}&\textbf{Algorithm}&\textbf{Category}&\textbf{A@1}&\textbf{A@1$\Uparrow$}&\textbf{A@2}&\textbf{A@2$\Uparrow$}&\textbf{A@3}&\textbf{A@3$\Uparrow$}&\textbf{Kappa Analysis} \\
\hline
         &\textbf{SLIM}& & 0.791&--&0.837&-- &0.86&--&0.7649 \\
         &DecisionTree& &0.532&48.7\% &0.635&31.8\% &0.656&31.1\%&0.5019 \\
         &Seer& &0.482&64.1\% &0.594&40.9\% &0.643&33.7\%&0.4507 \\
$\mathcal{A}$&MEPFL(RF)&Supervised& 0.698&13.3\% &0.837&0\% &0.86& 0\%& 0.7047 \\
      &MEFPL(MLP)& & 0.452 &75\% &0.574 &45.8\% &0.603 & 42.6\%& 0.3281\\
        &Dejavu&& 0.771&2.6\%&0.903& $-7.3$\%&0.934& $-7.9$\%& 0.7604\\
        &Eadro&& 0.741&6.7\%&0.86& $-2.7$\%&0.903& $-4.8$\%& 0.7214\\
        \cline{2-10}
        & AutoMap &Unsupervied& 0.336&135\%&0.435&92.4\%&0.489& 75.6\%&0.2745\\
        \cline{2-10}
        &Sage&Semi-supervised& 0.635&2.6\%&0.771& 8.6\%&0.837& 2.7\%& 0.6079\\
        &Murphy&& 0.656&2.6\%&0.791& 5.8\%&0.837& 2.7\%& 0.6242\\
\hline
\hline
         &\textbf{SLIM} && 0.673&--&0.75&--&0.827&--&0.6423\\
         &DecisionTree && 0.559&20.4\%&0.603&24.4\%&0.635& 30.2\%&0.5331\\
         &Seer && 0.503&33.8\%&0.564&33\%&0.603& 37.1\%&0.4882\\
$\mathcal{B}$&MEPFL(RF) &Supervised& 0.603&11.6\%&0.635&5.1\%&0.756& 9.4\%&0.5803\\
      &MEFPL(MLP) && 0.487&38.2\%&0.513&46.2\%&0.603& 37.1\%&0.4358\\
        &Dejavu&& 0.662&1.7\%&0.712&5.3\%&0.756& 9.4\%&0.6342\\
        &Eadro&& 0.635&6.0\%&0.698& 7.5\%&0.788& 4.5\%&0.6079 \\
        \cline{2-10}
        & AutoMap &Unsupervied& 0.258&161\%&0.342&119\%&0.379& 118\%&0.2132\\
        \cline{2-10}
        &Sage&Semi-supervised& 0.513&31.2\%&0.635& 18.1\%&0.712& 16.2\%&0.4570 \\
        &Murphy&& 0.564&19.3\%&0.662& 13.3\%&0.756& 9.4\%&0.5185 \\
\hline
\hline
         &\textbf{SLIM}&&0.931&--&0.967&--& 0.992 &--& 0.9132\\
         &DecisionTree && 0.771&20.8\%&0.86&12.4\%&0.90& 10.2\%&0.7332\\
         &Seer && 0.82&13.5\%&0.843&14.7\%&0.882&12.5\%&0.7913\\
$\mathcal{C}$&MEPFL(RF)&Supervised&  0.89&4.5\%&0.956&1.2\%&0.967& 2.6\%&0.8607\\
      &MEFPL(MLP) && 0.91&2.3\%&0.967&0\%&0.985&0.7\%&0.8764\\
        &Dejavu&& 0.92&0.4\%&0.956&1.1\%&0.992& 0\%&0.8832\\
        &Eadro&& 0.90&3.3\%&0.956& 1.1\%&0.992&  0\%& 0.8642\\
        \cline{2-10}
        & AutoMap &Unsupervied& 0.534&74.3\%&0.624&55\%&0.741&33.9\%&0.4213\\
        \cline{2-10}
        &Sage&Semi-supervised& 0.82&13.5\%&0.86& 8.2\%&0.90& 10.2\%&0.7862\\
        &Murphy&& 0.843&10.4\%&0.86& 12.4\%&0.90& 10.2\%&0.8135\\
\hline
\hline
\end{tabular}
\label{tab:acc}

\end{center}
\end{table*}

\begin{table}[t]
\caption{Accuracy comparison of different NumThresh}
\begin{center}
\begin{tabular}{p{0.9cm} p{1.6cm} p{0.6cm} p{0.6cm} p{0.6cm} p{0.6cm} p{0.6cm}}
\hline 
\textbf{Dataset}&\textbf{Bining Number}&\textbf{A@1}&\textbf{A@2}&\textbf{A@3}&\textbf{A@4}&\textbf{A@5}\\
\hline
         &50& 0.791& 0.814& 0.837&0.860& 0.860\\
         &75&0.791& 0.837&  0.837&  0.860& 0.860\\
$\mathcal{A}$&100&0.791& 0.837&  0.860& 0.884 &0.884\\
         &125&0.744&0.791&0.837&0.837&0.837\\
         &150&0.791& 0.837&  0.837&  0.837& 0.837\\
\hline
\hline
         &50& 0.632& 0.673& 0.712& 0.788& 0.884\\
         &75&0.632& 0.673& 0.788& 0.827& 0.884\\
$\mathcal{B}$&100&0.673& 0.75&  0.827& 0.884 &0.967\\
         &125&0.673& 0.712&  0.788& 0.827 &0.827\\
         &150&0.673& 0.712&  0.827& 0.884 &0.884 \\
\hline
\hline
         &50&0.82& 0.86&  0.891& 0.967 &1\\
         &75&0.82& 0.86&  0.891& 1 &1\\
$\mathcal{C}$&100&0.931& 0.967&  0.992& 1 &1\\
         &125&0.931& 0.967&  0.967& 1 &1\\
         &150&0.891& 0.967&  0.992& 1 &1 \\
\hline
\hline
\end{tabular}
\label{tab:thres}
\vspace{-15pt}
\end{center}
\end{table}


\subsubsection{Baseline Algorithms}  We compare our proposed SLIM
with several state-of-the-art fault localization algorithms, including five supervised methods, i.e. Dejavu, Seer, MEPFL-RandomForest(MEPFL-RF), Multilayer Perceptron(MEPFL-MLP), decision tree, Eadro, Murphy, Sage and AutoMap.
Dejavu is an actionable and interpretable fault localization method for recurring failures, where graph attention networks
is used to localize the fault~\cite{dejavu}. 
Murphy~\cite{murphy} based on a Markov Random Field (MRF) that can take advantage of such loose associations to reason about how entities affect each other in the context of a specific incident. 
Sage~\cite{sage} based on the Conditonal VAE that could simulate the service's status and counterfactual the system by restore the service's abnormal metric and confirm the root cause.
Eadro~\cite{eadro} is similar with Dejavu that leverages the Graph Attention Networks to learn the log, metric and trace. They try to embedding log into the node features by 
Seer~\cite{seer} captures the RPC-level graph dependency and metric by training a hybrid deep learning network that combines a CNN (Convolutional Neural Network) with an LSTM (Long Short-Term Memory). MEPFL-RandomForest(MEPFL-RF) and Multilayer Perceptron(MEPFL-MLP)~\cite{MEPFL} treat fault localization as a classification problem and solve it using traditional machine learning methods RandomForest and Multilayer Perceptron, respectively. We also compare our method with decision tree due to its interpretablity. AutoMap leverages the multi-dimension metrics to dynamically generate service relationship graph, and then leverages the random walk algorithm to localize the fault from the graph~\cite{automap}.

\subsubsection{Experiment Environment and Parameter Tuning} We implement SLIM using Python 3.7 and Go language. All the experiments are conducted on a personal computer with 3070ti, 32GB RAM and 5800X processors with 6 cores. We tune parameters for all methods by 5-fold cross-validation. Specifically, for our model SLIM, we choose the number of rules, i.e. parameter $K$, from $\{2, 4, 8, 12\}$, and limit the length of a rule to no more than $6$ to ensure interpretability, i.e., $l=6$. At the table~\ref{tab:thres}, we show the detail analysis of Feature Binarizer module. We test the performance impact of our rule set algorithm under different interval partition quantities and ultimately determined that the default optimal value is ~\textbf{100} for the partition quantity of Feature Binarizer.
For MEPFL-MLP and Seer, we adjust the number of neurons from [20,30,40] and the learning rate from [1e-4,5e-5,1e-5]. 
For DecisionTree and MEPFL(RF), the number of samples at each leaf node is tuned from 1 to 100 and the number of trees is tuned in $\{1000,2000,3000\}$.
For Dejavu, we set the parameters according to the suggestion in~\cite{dejavu}.

\subsection{Performance on Fault Localization}

\subsubsection{Evaluation Metrics}
\textbf{Top-k Accuracy}, which is referred as \textbf{A@k},
is used to measurement the perforemance of each methods.
Top-k Accuracy computes the probability that the root causes can be located within the top $k$ service instances among all candidates. Higher A@k indicates more accurate of the root cause localization. Here we measure the
performance of each method using $A@1$, $A@2$, and $A@3$.

\textbf{Kappa Analysis}, which is Cohen-Kappa analysis~\cite{cohenkappa}, a statistical method used to measure the inter-rater reliability. It is generally thought to be a more robust measure than a simple percent agreement calculation. Due to the requirement for precision data in Kappa analysis, and considering that our model provides root cause rankings rather than precision, we select the top-1(A@1) result from the ranking as the final localization outcome to carry out the Kappa test.

\subsubsection{Performance}
We present the fault localization results in table~\ref{tab:acc}, where the result are averaged over 5 independent trials. 
From table~\ref{tab:acc}, we see that the proposed SLIM achieves highest Top-1 accuracy and Cohen-Kappa value on dataset $\mathcal{A}$,$\mathcal{B}$ and $\mathcal{C}$. This due to we employ F1 score as objective function, which is robust to
data imbalance. In contrast, Dejavu simply resamples the data to balance the number of data in each class, thus performs slightly worse than SLIM~\cite{dejavu}. Eadro and Dejavu share similar performance outcomes because they employ the same methodology. Sage and murphy leverage the counterfactual method to restore the system and localize the root cause. Due to the Sage and Murphy is semi-supervised counterfactual inference methods, their performance experiences a slight decrease compared to supervised algorithms. However, they are more suitable for fault recovery and exploration in service change. Seer, Decision Tree, MEPFL-RF, and MEPFL-MLP do not have specific designs to address imbalanced data, leading to inferior performance. 

AutoMap performs poorly on three datasets. 
This is because these approaches do not make use of any historical faults information until the ground truth of similar historical faults are identified. 
Some critical intermediate steps in these methods,
such as anomaly detection and similarity evaluation,
despite being carefully designed, 
are entirely unsupervised. As a result, they may be susceptible to confusion from irrelevant abnormal changes in other metrics, which can be caused by noise or fluctuations, particularly when the number of metrics or fault units is high. On the contrary, SLIM focuses on the key metrics from the rule set that is generated by historical failures. 

\subsubsection{Overhead}
\begin{figure}
    \centering
\includegraphics[width=1\linewidth]{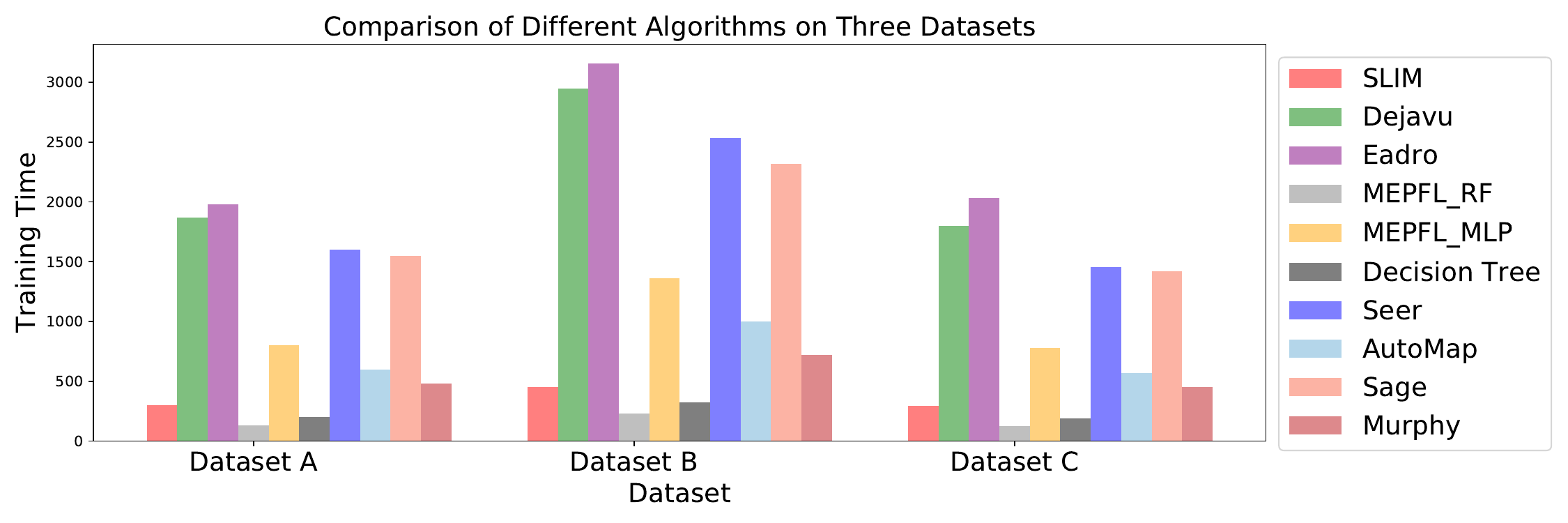}
    \caption{The Overhead of all Algorithms on Benchmark Datasets.}
    \label{fig:overhead1}
    \vspace{-15pt}
\end{figure}

In Figure~\ref{fig:overhead1}, we evaluate the training overhead for each algorithms on three datasets. From Figure~\ref{fig:overhead1}, we can see that our approach has lower training costs compared to all deep learning and some of machine learning models. 
As the deep learning based methods require more training time, thus
the computational costs of Seer and Dejavu
are much higher than that of the rest methods.
We note that as Seer and Dejavu are highly rely on the current service topology diagram, as 
the models need to be retained
once the service topology is change (such as there is a new service deployed).
Overall, the high training
overhead makes Seer and Dejavu unsuitable for scenarios
that needs the fault localization methods to be adapt quickly.


\subsection{Evaluation of Real-World System(Dataset D)}\label{csc}
We present the fault localization results in table~\ref{tab:acc_D}. Due to the large number of real system services and the high requirements for algorithmic overhead, the comparison algorithms we previously used, such as Dejavu and Seer, have excessive computational cost for deep learning models and cannot adapt well to the system requirements. Therefore, we only compare methods like Decision Tree and RandomForest. As shown in the table, our model still maintains better performance than other methods. In addition, we also compared the algorithm Ripper~\cite{cohen1995fast}, which is another rule-based algorithm, but it does not optimize for imbalanced data. Compared with Ripper, we made for the imbalanced dataset, our model has fewer false positives, resulting in a more accurate ranking of fault.

\begin{table}[htbp]
\caption{Comparison of different root cause localization algorithms at Real-World System.}
\begin{center}
\begin{tabular}{p{1cm} p{1.6cm} p{0.6cm} p{0.6cm} p{0.6cm} p{0.6cm} p{0.6cm} p{0.6cm}}
\hline 
\textbf{Dataset}&\textbf{Algorithm}&\textbf{A@1}&\textbf{A@1$\Uparrow$}&\textbf{A@2}&\textbf{A@2$\Uparrow$}&\textbf{A@3}&\textbf{A@3$\Uparrow$} \\
\hline
         &\textbf{SLIM} & 0.851&--&0.917&-- &0.965&-- \\
         $\mathcal{D}$&DecisionTree &0.742&14.7\% &0.832&10.2\% &0.88&10.0\% \\
         &MEPFL(RF)& 0.797&6.8\% &0.856&7.1\% &0.912& 5.8\% \\
        &Ripper& 0.723&17.7\%&0.813& 12.8\%&0.856& 12.7\% \\
\hline
\hline
\end{tabular}
\label{tab:acc_D}
\vspace{-15pt}
\end{center}
\end{table}




\subsection{Ability to Deal with Imbalanced Datasets}
\subsubsection{Experimental Setup}
To verify the performance advantages of our method in the data imbalance scenario, we extracted services to simulate the newly deployed services for the imbalance test. We selected the fault "network\_delay" in docker005(dataset $\mathcal{A}$), the fault "OS Network" in apache02(dataset $\mathcal{B}$) and the fault "OS Network" in Tomcat01(dataset $\mathcal{B}$). Because these faults occurs more frequently than others in the entire dataset, making it easier to characterize the trend of the fault localization performance with the frequency of occurrence. We set the test set for each service to include two faults, and the training set gradually increases from one occurrence to $n$-2, where $n$ is the total number of this faults.

\subsubsection{Numerical Results}\label{imb_exp}

\begin{figure}
    \centering
    \includegraphics[width=1\linewidth]{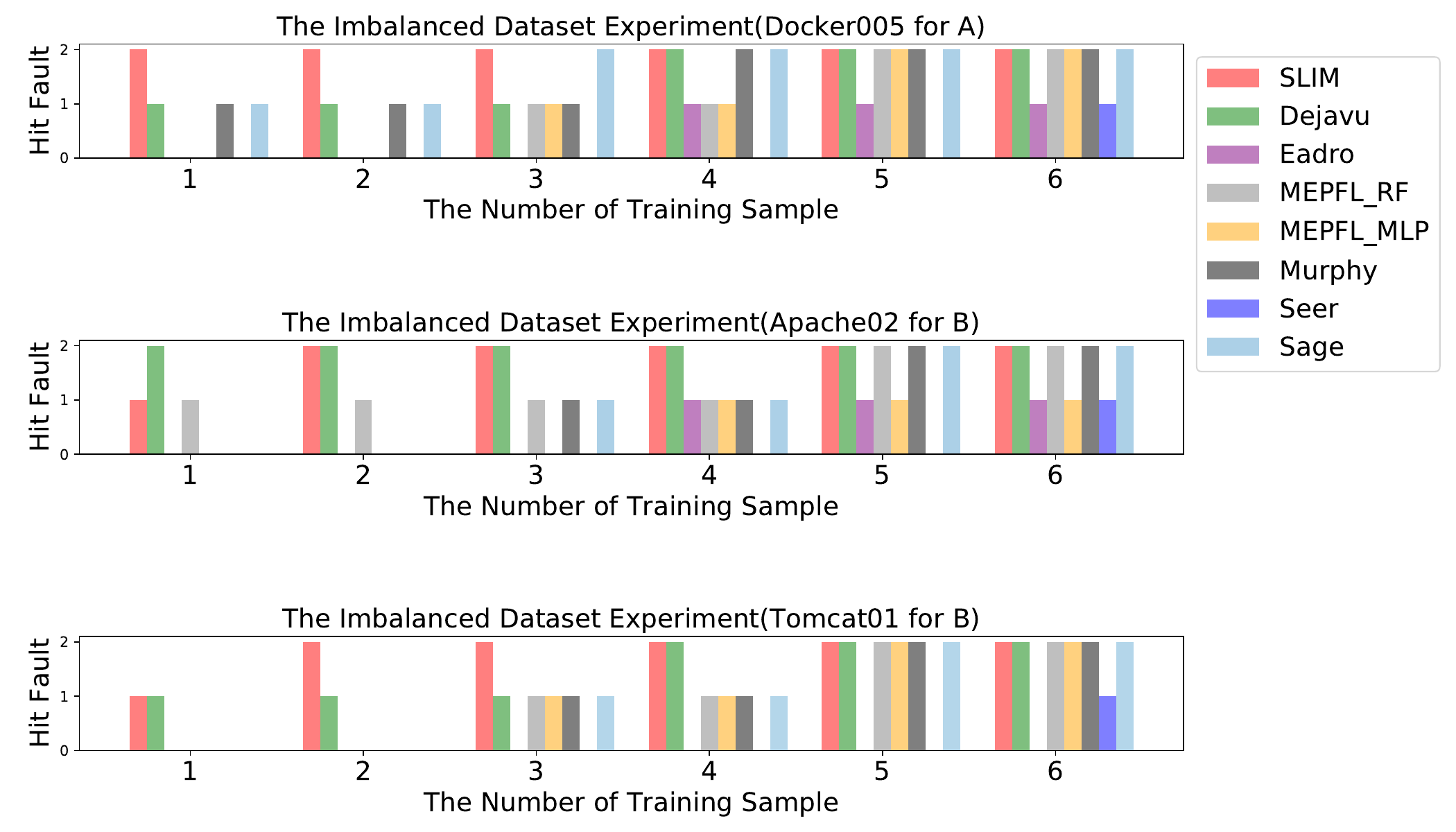}
    \caption{The Imbalanced Dataset Experiment.}
    \label{fig:imb_result}
    \vspace{-12pt}
\end{figure}

\begin{figure}
    \centering
    \includegraphics[width=1\linewidth]{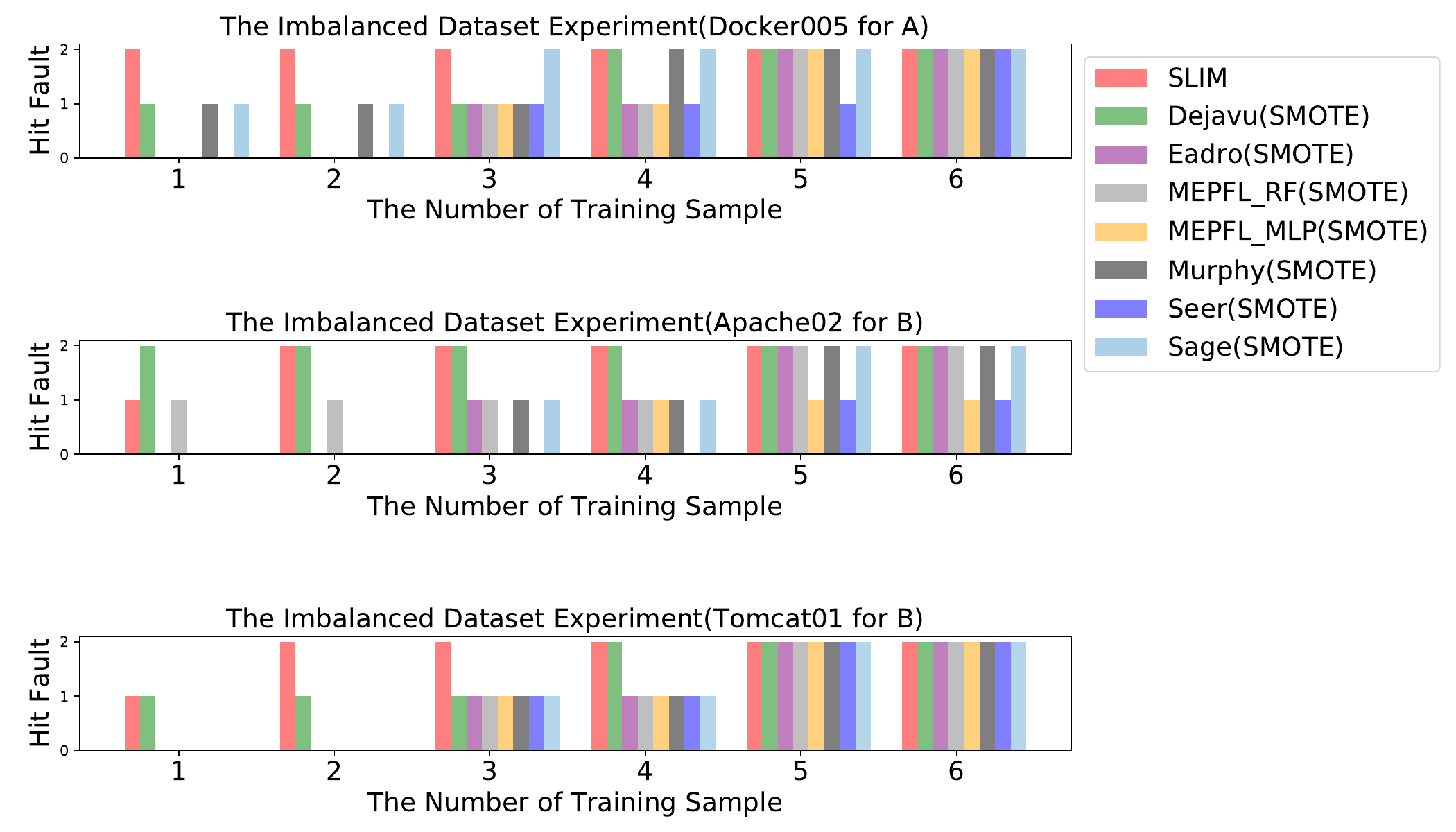}
    \caption{The Imbalanced Dataset Experiment with SMOTE.}
    \label{fig:imb_smote_result}
    \vspace{-15pt}
\end{figure}

We show the results of each fault localization algorithms
on three faults in Figure~\ref{fig:imb_result}. Noticed that in this experiment, we removed some underperforming algorithms from the previous performance comparison. Because we couldn't determine whether the poor performance of diagnosing new services was due to the inherent shortcomings of the models or if it was a result of imbalanced data.
From the Figure~\ref{fig:imb_result}, 
we find out that our model is able to correctly
identify all the testing faults 
when they occur in the training set for two times.
While the rest algorithms need more training
samples to produce reliable fault localization.
It means that, given a newly deployed service, 
our proposed SLIM needs only a few historical data to train,
thus can significantly reduce the numb fault the system needs to experience.
We further balance the number of faults by upsampling the data of minority faults using SMOTE~\cite{smote}.
We report the results of each method with SMOTE in Figure~\ref{fig:imb_smote_result}.
From Figure~\ref{fig:imb_smote_result}, we see that compared with the results in Figure~\ref{fig:imb_result}, few improvement is achieved when SMOTE is used
for most algorithms.
In Apache02 for Dataset $\mathcal{B}$, we find out that the Seer has a little promotion. However, the SMOTE produced negative impacts for Seer on fault ``docker005''
in Dataset $\mathcal{A}$. This is due to that SMOTE may introduce some noisy data to the model, which may affect the precision of Seer. 

\subsection{Performance of Knowledge Base Generator}
We evaluate the generator using precision and time consumption. Two AIOPS experts are recruited to manually finish the knowledge base construction by their expert knowledge. The experts have longer than two years of operating experience and have publications at national conferences. As a baseline, we also construct the knowledge base using DecisionTree and Dejavu(local interpreter part). These are done using similar procedure to that used for SLIM.
Table \ref{tab:generator} shows the comparison results of our interpreter, expert knowledge and other baseline method for CloudRCA and MicroCBR using dataset A. Compared with the expert knowledge, our interpreter finishes the root cause analysis automatically and loses just 5.1\%--7.8\% precision. Compared with other baseline methods, our interpreter's knowledge base improves precision by 7.8\%-19.4\%.

We also compare the time consumption by all methods in dataset A. 
Table \ref{tab:generator} compares the results between the experts and the generator. In the experiment, experts require 5 hours to construct the knowledge base. Then, we compare the knowledge base effect among experts, our interpreter and other baseline interpreter. Compared with experts, our interpreter reduces the time required by 82\% and has nearly the same precision.

\begin{table}[htbp]
\vspace{-10pt}
\caption{The Performance Comparison of Generator and Expert (Pre:Precision; TC:time-consuming)}
\begin{center}
\begin{tabular}{ p{3.5cm} p{0.5cm} p{1cm} p{0.7cm} p{1cm}}
\hline
\textbf{Algorithm}&\textbf{Pre}&\textbf{Pre$\uparrow$}&\textbf{TC}&\textbf{TC$\downarrow$}\\
\hline
\textbf{CloudRCA(SLIM)}&70\%&--&57min&--\\
CloudRCA(DecisionTree)&60\%&16.7\%&46min&-24\%\\
CloudRCA(Dejavu)&64\%&7.8\%&89min&36\%\\
CloudRCA(Expert)&76\%&-7.8\%&5h&82\%\\
\hline
\textbf{MicroCBR(Interpreter)}&74\%&--\%&55min&--\\
MicroCBR(DecisionTree)&62\%&19.4\%&43min&-28\%\\
MicroCBR(Dejavu)&66\%&11\%&87&36.7\%\\
MicroCBR(Expert)&78\%&5.1\%&5h&82\%\\
\hline
\end{tabular}
\label{tab:generator}
\end{center}
\vspace{-22pt}
\end{table}

\section{Limitation and Futurework}
Due to the complexity of F1-Score in multi-class settings, our model, in order to trade off computational cost and performance, is optimized only for binary classification. This design choice makes our model not strictly end-to-end, which may decrease performance in the ultimate fault localization. 

In future work, we aim to propose a new multi-class F1-Score optimization method to learn rule sets based on this module, achieving a fully end-to-end model. This approach seeks to address the performance trade-off issues encountered in the previous implementation.
\section{Conclusion}

In this paper, we propose an interpretable, effective and fast fault localization algorithm SLIM to directly optimize the F1 score, which is particularly applicable for highly imbalanced classification. Our experimental results demonstrate the superior performance and interpretability of SLIM in comparison with existing fault localization methods. In addition, the good adaptbility of SLIM makes it an ideal tool to handle large-scale microservice systems in many real-world scenarios involving frequent service change.

\bibliographystyle{ACM-Reference-Format}
\bibliography{bib}


\begin{thebibliography}{42}


\ifx \showCODEN    \undefined \def \showCODEN     #1{\unskip}     \fi
\ifx \showDOI      \undefined \def \showDOI       #1{#1}\fi
\ifx \showISBNx    \undefined \def \showISBNx     #1{\unskip}     \fi
\ifx \showISBNxiii \undefined \def \showISBNxiii  #1{\unskip}     \fi
\ifx \showISSN     \undefined \def \showISSN      #1{\unskip}     \fi
\ifx \showLCCN     \undefined \def \showLCCN      #1{\unskip}     \fi
\ifx \shownote     \undefined \def \shownote      #1{#1}          \fi
\ifx \showarticletitle \undefined \def \showarticletitle #1{#1}   \fi
\ifx \showURL      \undefined \def \showURL       {\relax}        \fi
\providecommand\bibfield[2]{#2}
\providecommand\bibinfo[2]{#2}
\providecommand\natexlab[1]{#1}
\providecommand\showeprint[2][]{arXiv:#2}

\bibitem[\protect\citeauthoryear{??}{dat}{2023a}]%
        {datasetAB}
 \bibinfo{year}{2023}\natexlab{a}.
\newblock \bibinfo{title}{Dataset A and B.}
\newblock \bibinfo{howpublished}{\url{https://github.com/NetManAIOps/DejaVu}}.   (\bibinfo{year}{2023}).
\newblock


\bibitem[\protect\citeauthoryear{??}{dat}{2023b}]%
        {datasetC}
 \bibinfo{year}{2023}\natexlab{b}.
\newblock \bibinfo{title}{Dataset C.}
\newblock \bibinfo{howpublished}{\url{https://github.com/NetManAIOps/TraceRCA}}.   (\bibinfo{year}{2023}).
\newblock


\bibitem[\protect\citeauthoryear{??}{ver}{2023}]%
        {ver2}
 \bibinfo{year}{2023}\natexlab{}.
\newblock \bibinfo{title}{Microsoft Doc}.
\newblock \bibinfo{howpublished}{\url{https://learn.microsoft.com/en-us/azure/architecture/guide/architecture-styles/microservices}}.   (\bibinfo{year}{2023}).
\newblock


\bibitem[\protect\citeauthoryear{Ahmed, Ghosh, Bansal, Zimmermann, Zhang, and Rajmohan}{Ahmed et~al\mbox{.}}{2023}]%
        {LLM}
\bibfield{author}{\bibinfo{person}{Toufique Ahmed}, \bibinfo{person}{Supriyo Ghosh}, \bibinfo{person}{Chetan Bansal}, \bibinfo{person}{Thomas Zimmermann}, \bibinfo{person}{Xuchao Zhang}, {and} \bibinfo{person}{Saravan Rajmohan}.} \bibinfo{year}{2023}\natexlab{}.
\newblock \showarticletitle{Recommending Root-Cause and Mitigation Steps for Cloud Incidents using Large Language Models}.
\newblock \bibinfo{journal}{{\em arXiv preprint arXiv:2301.03797\/}} (\bibinfo{year}{2023}).
\newblock


\bibitem[\protect\citeauthoryear{Bach et~al\mbox{.}}{Bach et~al\mbox{.}}{2013}]%
        {bach2013learning}
\bibfield{author}{\bibinfo{person}{Francis Bach} {et~al\mbox{.}}} \bibinfo{year}{2013}\natexlab{}.
\newblock \showarticletitle{Learning with submodular functions: A convex optimization perspective}.
\newblock \bibinfo{journal}{{\em Foundations and Trends{\textregistered} in Machine Learning\/}} \bibinfo{volume}{6}, \bibinfo{number}{2-3} (\bibinfo{year}{2013}), \bibinfo{pages}{145--373}.
\newblock


\bibitem[\protect\citeauthoryear{Beyer, Jones, Petoff, and Murphy}{Beyer et~al\mbox{.}}{2016}]%
        {ver3}
\bibfield{author}{\bibinfo{person}{Betsy Beyer}, \bibinfo{person}{Chris Jones}, \bibinfo{person}{Jennifer Petoff}, {and} \bibinfo{person}{Niall~Richard Murphy}.} \bibinfo{year}{2016}\natexlab{}.
\newblock \bibinfo{booktitle}{{\em Site Reliability Engineering: How Google Runs Production Systems}}.
\newblock
\showURL{%
\url{http://landing.google.com/sre/book.html}}


\bibitem[\protect\citeauthoryear{Brand{\'o}n, Sol{\'e}, Hu{\'e}lamo, Solans, P{\'e}rez, and Munt{\'e}s-Mulero}{Brand{\'o}n et~al\mbox{.}}{2020}]%
        {intro1}
\bibfield{author}{\bibinfo{person}{{\'A}lvaro Brand{\'o}n}, \bibinfo{person}{Marc Sol{\'e}}, \bibinfo{person}{Alberto Hu{\'e}lamo}, \bibinfo{person}{David Solans}, \bibinfo{person}{Mar{\'\i}a~S P{\'e}rez}, {and} \bibinfo{person}{Victor Munt{\'e}s-Mulero}.} \bibinfo{year}{2020}\natexlab{}.
\newblock \showarticletitle{Graph-based root cause analysis for service-oriented and microservice architectures}.
\newblock \bibinfo{journal}{{\em Journal of Systems and Software\/}}  \bibinfo{volume}{159} (\bibinfo{year}{2020}), \bibinfo{pages}{110432}.
\newblock


\bibitem[\protect\citeauthoryear{Chawla, Bowyer, Hall, and Kegelmeyer}{Chawla et~al\mbox{.}}{2002}]%
        {smote}
\bibfield{author}{\bibinfo{person}{Nitesh~V Chawla}, \bibinfo{person}{Kevin~W Bowyer}, \bibinfo{person}{Lawrence~O Hall}, {and} \bibinfo{person}{W~Philip Kegelmeyer}.} \bibinfo{year}{2002}\natexlab{}.
\newblock \showarticletitle{SMOTE: synthetic minority over-sampling technique}.
\newblock \bibinfo{journal}{{\em Journal of artificial intelligence research\/}}  \bibinfo{volume}{16} (\bibinfo{year}{2002}), \bibinfo{pages}{321--357}.
\newblock


\bibitem[\protect\citeauthoryear{Chen, He, Lin, Xu, Zhang, Hao, Gao, Xu, Dang, and Zhang}{Chen et~al\mbox{.}}{2019a}]%
        {icse7}
\bibfield{author}{\bibinfo{person}{Junjie Chen}, \bibinfo{person}{Xiaoting He}, \bibinfo{person}{Qingwei Lin}, \bibinfo{person}{Yong Xu}, \bibinfo{person}{Hongyu Zhang}, \bibinfo{person}{Dan Hao}, \bibinfo{person}{Feng Gao}, \bibinfo{person}{Zhangwei Xu}, \bibinfo{person}{Yingnong Dang}, {and} \bibinfo{person}{Dongmei Zhang}.} \bibinfo{year}{2019}\natexlab{a}.
\newblock \showarticletitle{An empirical investigation of incident triage for online service systems}. In \bibinfo{booktitle}{{\em 2019 IEEE/ACM 41st International Conference on Software Engineering: Software Engineering in Practice (ICSE-SEIP)}}. IEEE, \bibinfo{pages}{111--120}.
\newblock


\bibitem[\protect\citeauthoryear{Chen, He, Lin, Zhang, Hao, Gao, Xu, Dang, and Zhang}{Chen et~al\mbox{.}}{2019b}]%
        {icse6}
\bibfield{author}{\bibinfo{person}{Junjie Chen}, \bibinfo{person}{Xiaoting He}, \bibinfo{person}{Qingwei Lin}, \bibinfo{person}{Hongyu Zhang}, \bibinfo{person}{Dan Hao}, \bibinfo{person}{Feng Gao}, \bibinfo{person}{Zhangwei Xu}, \bibinfo{person}{Yingnong Dang}, {and} \bibinfo{person}{Dongmei Zhang}.} \bibinfo{year}{2019}\natexlab{b}.
\newblock \showarticletitle{Continuous incident triage for large-scale online service systems}. In \bibinfo{booktitle}{{\em 2019 34th IEEE/ACM International Conference on Automated Software Engineering (ASE)}}. IEEE, \bibinfo{pages}{364--375}.
\newblock


\bibitem[\protect\citeauthoryear{Chen, Ouyang, and Zhang}{Chen et~al\mbox{.}}{2021}]%
        {icse1}
\bibfield{author}{\bibinfo{person}{Lingchao Chen}, \bibinfo{person}{Yicheng Ouyang}, {and} \bibinfo{person}{Lingming Zhang}.} \bibinfo{year}{2021}\natexlab{}.
\newblock \showarticletitle{Fast and precise on-the-fly patch validation for all}. In \bibinfo{booktitle}{{\em 2021 IEEE/ACM 43rd International Conference on Software Engineering (ICSE)}}. IEEE, \bibinfo{pages}{1123--1134}.
\newblock


\bibitem[\protect\citeauthoryear{Cohen}{Cohen}{1995}]%
        {cohen1995fast}
\bibfield{author}{\bibinfo{person}{William~W Cohen}.} \bibinfo{year}{1995}\natexlab{}.
\newblock \showarticletitle{Fast effective rule induction}.
\newblock In \bibinfo{booktitle}{{\em Machine learning proceedings 1995}}. \bibinfo{publisher}{Elsevier}.
\newblock


\bibitem[\protect\citeauthoryear{Dash, Gunluk, and Wei}{Dash et~al\mbox{.}}{2018}]%
        {CG-Dash}
\bibfield{author}{\bibinfo{person}{Sanjeeb Dash}, \bibinfo{person}{Oktay Gunluk}, {and} \bibinfo{person}{Dennis Wei}.} \bibinfo{year}{2018}\natexlab{}.
\newblock \showarticletitle{Boolean Decision Rules via Column Generation}. In \bibinfo{booktitle}{{\em Advances in Neural Information Processing Systems}}.
\newblock


\bibitem[\protect\citeauthoryear{Fengrui~L}{Fengrui~L}{2022}]%
        {microcbr}
\bibfield{author}{\bibinfo{person}{Yang~W Fengrui~L}.} \bibinfo{year}{2022}\natexlab{}.
\newblock \showarticletitle{MicroCBR: Case-based Reasoning on Spatio-temporal Fault Knowledge Graph for Microservices Troubleshooting}. In \bibinfo{booktitle}{{\em International Conference on Case-Based Reasoning}}.
\newblock


\bibitem[\protect\citeauthoryear{Gan, Liang, et~al\mbox{.}}{Gan et~al\mbox{.}}{2021}]%
        {sage}
\bibfield{author}{\bibinfo{person}{Yu Gan}, \bibinfo{person}{Mingyu Liang}, {et~al\mbox{.}}} \bibinfo{year}{2021}\natexlab{}.
\newblock \showarticletitle{Sage: Using Unsupervised Learning for Scalable Performance Debugging in Microservices}.
\newblock \bibinfo{journal}{{\em arXiv preprint arXiv:2101.00267\/}} (\bibinfo{year}{2021}).
\newblock


\bibitem[\protect\citeauthoryear{Gan, Zhang, et~al\mbox{.}}{Gan et~al\mbox{.}}{2019a}]%
        {deathstar}
\bibfield{author}{\bibinfo{person}{Yu Gan}, \bibinfo{person}{Yanqi Zhang}, {et~al\mbox{.}}} \bibinfo{year}{2019}\natexlab{a}.
\newblock \showarticletitle{An Open-Source Benchmark Suite for Microservices and Their Hardware-Software Implications for Cloud {\&} Edge Systems}. In \bibinfo{booktitle}{{\em Proceedings of the Twenty-Fourth International Conference on Architectural Support for Programming Languages and Operating Systems, {ASPLOS} 2019, Providence, RI, USA, April 13-17, 2019}}, \bibfield{editor}{\bibinfo{person}{Iris Bahar}, \bibinfo{person}{Maurice Herlihy}, \bibinfo{person}{Emmett Witchel}, {and} \bibinfo{person}{Alvin~R. Lebeck}} (Eds.). \bibinfo{publisher}{{ACM}}, \bibinfo{pages}{3--18}.
\newblock
\showDOI{%
\url{https://doi.org/10.1145/3297858.3304013}}


\bibitem[\protect\citeauthoryear{Gan, Zhang, et~al\mbox{.}}{Gan et~al\mbox{.}}{2019b}]%
        {seer}
\bibfield{author}{\bibinfo{person}{Yu Gan}, \bibinfo{person}{Yanqi Zhang}, {et~al\mbox{.}}} \bibinfo{year}{2019}\natexlab{b}.
\newblock \showarticletitle{Seer: Leveraging big data to navigate the complexity of performance debugging in cloud microservices}. In \bibinfo{booktitle}{{\em Proceedings of the twenty-fourth international conference on architectural support for programming languages and operating systems}}. \bibinfo{pages}{19--33}.
\newblock


\bibitem[\protect\citeauthoryear{Harsh, Zhou, Ashok, Mysore, Godfrey, and Banerjee}{Harsh et~al\mbox{.}}{2023}]%
        {murphy}
\bibfield{author}{\bibinfo{person}{Vipul Harsh}, \bibinfo{person}{Wenxuan Zhou}, \bibinfo{person}{Sachin Ashok}, \bibinfo{person}{Radhika~Niranjan Mysore}, \bibinfo{person}{Brighten Godfrey}, {and} \bibinfo{person}{Sujata Banerjee}.} \bibinfo{year}{2023}\natexlab{}.
\newblock \showarticletitle{Murphy: Performance Diagnosis of Distributed Cloud Applications}. In \bibinfo{booktitle}{{\em Proceedings of the ACM SIGCOMM 2023 Conference}}. \bibinfo{pages}{438--451}.
\newblock


\bibitem[\protect\citeauthoryear{Harshaw, Feldman, Ward, and Karbasi}{Harshaw et~al\mbox{.}}{2019}]%
        {harshaw2019submodular}
\bibfield{author}{\bibinfo{person}{Chris Harshaw}, \bibinfo{person}{Moran Feldman}, \bibinfo{person}{Justin Ward}, {and} \bibinfo{person}{Amin Karbasi}.} \bibinfo{year}{2019}\natexlab{}.
\newblock \showarticletitle{Submodular Maximization beyond Non-negativity: Guarantees, Fast Algorithms, and Applications}. In \bibinfo{booktitle}{{\em Proceedings of the 36th ICML}}, Vol.~\bibinfo{volume}{97}. \bibinfo{publisher}{PMLR}, \bibinfo{pages}{2634--2643}.
\newblock


\bibitem[\protect\citeauthoryear{He, Zhu, and Zheng}{He et~al\mbox{.}}{2017}]%
        {drain}
\bibfield{author}{\bibinfo{person}{Pinjia He}, \bibinfo{person}{Jieming Zhu}, {and} \bibinfo{person}{Zibin Zheng}.} \bibinfo{year}{2017}\natexlab{}.
\newblock \showarticletitle{Drain: An online log parsing approach with fixed depth tree}. In \bibinfo{booktitle}{{\em 2017 IEEE international conference on web services (ICWS)}}. IEEE, \bibinfo{pages}{33--40}.
\newblock


\bibitem[\protect\citeauthoryear{Hunter and Lange}{Hunter and Lange}{2004}]%
        {hunter2004tutorial}
\bibfield{author}{\bibinfo{person}{David~R Hunter} {and} \bibinfo{person}{Kenneth Lange}.} \bibinfo{year}{2004}\natexlab{}.
\newblock \showarticletitle{A tutorial on MM algorithms}.
\newblock \bibinfo{journal}{{\em The American Statistician\/}}  \bibinfo{volume}{58} (\bibinfo{year}{2004}).
\newblock


\bibitem[\protect\citeauthoryear{Jegelka and Bilmes}{Jegelka and Bilmes}{2011}]%
        {jegelka2011submodularity}
\bibfield{author}{\bibinfo{person}{S Jegelka} {and} \bibinfo{person}{J Bilmes}.} \bibinfo{year}{2011}\natexlab{}.
\newblock \showarticletitle{Submodularity beyond submodular energies: Coupling edges in graph cuts}. In \bibinfo{booktitle}{{\em Proceedings of the 2011 IEEE Conference on CVPR}}.
\newblock


\bibitem[\protect\citeauthoryear{Jiang, Liu, Niu, Zhang, and Hu}{Jiang et~al\mbox{.}}{2021}]%
        {icse2}
\bibfield{author}{\bibinfo{person}{Yanjie Jiang}, \bibinfo{person}{Hui Liu}, \bibinfo{person}{Nan Niu}, \bibinfo{person}{Lu Zhang}, {and} \bibinfo{person}{Yamin Hu}.} \bibinfo{year}{2021}\natexlab{}.
\newblock \showarticletitle{Extracting concise bug-fixing patches from human-written patches in version control systems}. In \bibinfo{booktitle}{{\em 2021 IEEE/ACM 43rd International Conference on Software Engineering (ICSE)}}. IEEE, \bibinfo{pages}{686--698}.
\newblock


\bibitem[\protect\citeauthoryear{Johnson and Khoshgoftaar}{Johnson and Khoshgoftaar}{2019}]%
        {Johnson2019:survey}
\bibfield{author}{\bibinfo{person}{Justin~M. Johnson} {and} \bibinfo{person}{Taghi~M. Khoshgoftaar}.} \bibinfo{year}{2019}\natexlab{}.
\newblock \showarticletitle{Survey on Deep Learning with Class Imbalance}.
\newblock \bibinfo{journal}{{\em Journal of Big Data\/}} \bibinfo{volume}{6}, \bibinfo{number}{27} (\bibinfo{year}{2019}).
\newblock


\bibitem[\protect\citeauthoryear{Lee, Yang, Chen, Su, and Lyu}{Lee et~al\mbox{.}}{2023}]%
        {eadro}
\bibfield{author}{\bibinfo{person}{Cheryl Lee}, \bibinfo{person}{Tianyi Yang}, \bibinfo{person}{Zhuangbin Chen}, \bibinfo{person}{Yuxin Su}, {and} \bibinfo{person}{Michael~R Lyu}.} \bibinfo{year}{2023}\natexlab{}.
\newblock \showarticletitle{Eadro: An End-to-End Troubleshooting Framework for Microservices on Multi-source Data}.
\newblock \bibinfo{journal}{{\em arXiv preprint arXiv:2302.05092\/}} (\bibinfo{year}{2023}).
\newblock


\bibitem[\protect\citeauthoryear{Lee, Mirrokni, Nagarajan, and Sviridenko}{Lee et~al\mbox{.}}{2010}]%
        {lee2010maximizing}
\bibfield{author}{\bibinfo{person}{Jon Lee}, \bibinfo{person}{Vahab~S Mirrokni}, \bibinfo{person}{Viswanath Nagarajan}, {and} \bibinfo{person}{Maxim Sviridenko}.} \bibinfo{year}{2010}\natexlab{}.
\newblock \showarticletitle{Maximizing nonmonotone submodular functions under matroid or knapsack constraints}.
\newblock \bibinfo{journal}{{\em SIAM Journal on Discrete Mathematics\/}} \bibinfo{volume}{23}, \bibinfo{number}{4} (\bibinfo{year}{2010}).
\newblock


\bibitem[\protect\citeauthoryear{Li, Chen, Lin, Wang, and Chen}{Li et~al\mbox{.}}{2021b}]%
        {ver1}
\bibfield{author}{\bibinfo{person}{Xing Li}, \bibinfo{person}{Yan Chen}, \bibinfo{person}{Zhiqiang Lin}, \bibinfo{person}{Xiao Wang}, {and} \bibinfo{person}{Jim~Hao Chen}.} \bibinfo{year}{2021}\natexlab{b}.
\newblock \showarticletitle{Automatic policy generation for $\{$Inter-Service$\}$ access control of microservices}. In \bibinfo{booktitle}{{\em 30th USENIX Security Symposium (USENIX Security 21)}}. \bibinfo{pages}{3971--3988}.
\newblock


\bibitem[\protect\citeauthoryear{Li, Chen, et~al\mbox{.}}{Li et~al\mbox{.}}{2021a}]%
        {tracerca}
\bibfield{author}{\bibinfo{person}{Zeyan Li}, \bibinfo{person}{Junjie Chen}, {et~al\mbox{.}}} \bibinfo{year}{2021}\natexlab{a}.
\newblock \showarticletitle{Practical root cause localization for microservice systems via trace analysis}. In \bibinfo{booktitle}{{\em 2021 IEEE/ACM 29th International Symposium on Quality of Service (IWQOS)}}. IEEE, \bibinfo{pages}{1--10}.
\newblock


\bibitem[\protect\citeauthoryear{Li, Zhao, Li, Lu, Wang, Chang, Nie, Cao, Zhang, Sui, Wang, Du, Duan, and Pei}{Li et~al\mbox{.}}{2022}]%
        {dejavu}
\bibfield{author}{\bibinfo{person}{Zeyan Li}, \bibinfo{person}{Nengwen Zhao}, \bibinfo{person}{Mingjie Li}, \bibinfo{person}{Xianglin Lu}, \bibinfo{person}{Lixin Wang}, \bibinfo{person}{Dongdong Chang}, \bibinfo{person}{Xiaohui Nie}, \bibinfo{person}{Li Cao}, \bibinfo{person}{Wenchi Zhang}, \bibinfo{person}{Kaixin Sui}, \bibinfo{person}{Yanhua Wang}, \bibinfo{person}{Xu Du}, \bibinfo{person}{Guoqing Duan}, {and} \bibinfo{person}{Dan Pei}.} \bibinfo{year}{2022}\natexlab{}.
\newblock \showarticletitle{Actionable and Interpretable Fault Localization for Recurring Failures in Online Service Systems}. In \bibinfo{booktitle}{{\em Proceedings of the 2022 30th {{ACM Joint Meeting}} on {{European Software Engineering Conference}} and {{Symposium}} on the {{Foundations}} of {{Software Engineering}}}} {\em (\bibinfo{series}{{{ESEC}}/{{FSE}} 2022})}.
\newblock


\bibitem[\protect\citeauthoryear{Lin, Zhang, Lou, Zhang, and Chen}{Lin et~al\mbox{.}}{2016}]%
        {intro2}
\bibfield{author}{\bibinfo{person}{Qingwei Lin}, \bibinfo{person}{Hongyu Zhang}, \bibinfo{person}{Jian-Guang Lou}, \bibinfo{person}{Yu Zhang}, {and} \bibinfo{person}{Xuewei Chen}.} \bibinfo{year}{2016}\natexlab{}.
\newblock \showarticletitle{Log clustering based problem identification for online service systems}. In \bibinfo{booktitle}{{\em Proceedings of the 38th International Conference on Software Engineering Companion}}. \bibinfo{pages}{102--111}.
\newblock


\bibitem[\protect\citeauthoryear{Ma, Xu, et~al\mbox{.}}{Ma et~al\mbox{.}}{2020a}]%
        {automap}
\bibfield{author}{\bibinfo{person}{Meng Ma}, \bibinfo{person}{Jingmin Xu}, {et~al\mbox{.}}} \bibinfo{year}{2020}\natexlab{a}.
\newblock \showarticletitle{Automap: Diagnose your microservice-based web applications automatically}. In \bibinfo{booktitle}{{\em Proceedings of The Web Conference 2020}}. \bibinfo{pages}{246--258}.
\newblock


\bibitem[\protect\citeauthoryear{Ma, Yin, Zhang, Wang, Zheng, Jiang, Hu, Luo, Li, Qiu, et~al\mbox{.}}{Ma et~al\mbox{.}}{2020b}]%
        {icse8}
\bibfield{author}{\bibinfo{person}{Minghua Ma}, \bibinfo{person}{Zheng Yin}, \bibinfo{person}{Shenglin Zhang}, \bibinfo{person}{Sheng Wang}, \bibinfo{person}{Christopher Zheng}, \bibinfo{person}{Xinhao Jiang}, \bibinfo{person}{Hanwen Hu}, \bibinfo{person}{Cheng Luo}, \bibinfo{person}{Yilin Li}, \bibinfo{person}{Nengjun Qiu}, {et~al\mbox{.}}} \bibinfo{year}{2020}\natexlab{b}.
\newblock \showarticletitle{Diagnosing root causes of intermittent slow queries in cloud databases}.
\newblock \bibinfo{journal}{{\em Proceedings of the VLDB Endowment\/}} \bibinfo{volume}{13}, \bibinfo{number}{8} (\bibinfo{year}{2020}), \bibinfo{pages}{1176--1189}.
\newblock


\bibitem[\protect\citeauthoryear{Mahimkar}{Mahimkar}{2011}]%
        {intro8}
\bibfield{author}{\bibinfo{person}{Ajay Mahimkar}.} \bibinfo{year}{2011}\natexlab{}.
\newblock \showarticletitle{Rapid detection of maintenance induced changes in service performance}. In \bibinfo{booktitle}{{\em Proceedings of the Seventh COnference on Emerging Networking EXperiments and Technologies}}. \bibinfo{pages}{1--12}.
\newblock


\bibitem[\protect\citeauthoryear{Mariani, Pezz{\`e}, Riganelli, and Xin}{Mariani et~al\mbox{.}}{2020}]%
        {intro3}
\bibfield{author}{\bibinfo{person}{Leonardo Mariani}, \bibinfo{person}{Mauro Pezz{\`e}}, \bibinfo{person}{Oliviero Riganelli}, {and} \bibinfo{person}{Rui Xin}.} \bibinfo{year}{2020}\natexlab{}.
\newblock \showarticletitle{Predicting failures in multi-tier distributed systems}.
\newblock \bibinfo{journal}{{\em Journal of Systems and Software\/}}  \bibinfo{volume}{161} (\bibinfo{year}{2020}), \bibinfo{pages}{110464}.
\newblock


\bibitem[\protect\citeauthoryear{McHugh}{McHugh}{2012}]%
        {cohenkappa}
\bibfield{author}{\bibinfo{person}{Mary~L McHugh}.} \bibinfo{year}{2012}\natexlab{}.
\newblock \showarticletitle{Interrater reliability: the kappa statistic}.
\newblock \bibinfo{journal}{{\em Biochemia medica\/}} \bibinfo{volume}{22}, \bibinfo{number}{3} (\bibinfo{year}{2012}), \bibinfo{pages}{276--282}.
\newblock


\bibitem[\protect\citeauthoryear{Mehta and Bhagwan}{Mehta and Bhagwan}{2020}]%
        {intro7}
\bibfield{author}{\bibinfo{person}{Sonu Mehta} {and} \bibinfo{person}{Ranjita Bhagwan}.} \bibinfo{year}{2020}\natexlab{}.
\newblock \showarticletitle{Rex: Preventing bugs and misconfiguration in large services using correlated change analysis}. In \bibinfo{booktitle}{{\em 17th USENIX Symposium on Networked Systems Design and Implementation (NSDI 20)}}. \bibinfo{pages}{435--448}.
\newblock


\bibitem[\protect\citeauthoryear{Nemhauser, Wolsey, and Fisher}{Nemhauser et~al\mbox{.}}{1978}]%
        {nemhauser1978analysis}
\bibfield{author}{\bibinfo{person}{George~L Nemhauser}, \bibinfo{person}{Laurence~A Wolsey}, {and} \bibinfo{person}{Marshall~L Fisher}.} \bibinfo{year}{1978}\natexlab{}.
\newblock \showarticletitle{An analysis of approximations for maximizing submodular set functions—{I}}.
\newblock \bibinfo{journal}{{\em Mathematical Programming\/}} \bibinfo{volume}{14}, \bibinfo{number}{1} (\bibinfo{year}{1978}), \bibinfo{pages}{265--294}.
\newblock


\bibitem[\protect\citeauthoryear{Sriraman and Wenisch}{Sriraman and Wenisch}{2018}]%
        {suite}
\bibfield{author}{\bibinfo{person}{Akshitha Sriraman} {and} \bibinfo{person}{Thomas~F Wenisch}.} \bibinfo{year}{2018}\natexlab{}.
\newblock \showarticletitle{$\mu$ suite: a benchmark suite for microservices}. In \bibinfo{booktitle}{{\em 2018 IEEE International Symposium on Workload Characterization (IISWC)}}. IEEE, \bibinfo{pages}{1--12}.
\newblock


\bibitem[\protect\citeauthoryear{Zhang, Guan, et~al\mbox{.}}{Zhang et~al\mbox{.}}{2021}]%
        {cloudrca}
\bibfield{author}{\bibinfo{person}{Yingying Zhang}, \bibinfo{person}{Zhengxiong Guan}, {et~al\mbox{.}}} \bibinfo{year}{2021}\natexlab{}.
\newblock \showarticletitle{CloudRCA: A root cause analysis framework for cloud computing platforms}. In \bibinfo{booktitle}{{\em Proceedings of the 30th ACM International Conference on Information \& Knowledge Management}}. \bibinfo{pages}{4373--4382}.
\newblock


\bibitem[\protect\citeauthoryear{Zhao, Chen, Yu, et~al\mbox{.}}{Zhao et~al\mbox{.}}{2021}]%
        {scwarn}
\bibfield{author}{\bibinfo{person}{Nengwen Zhao}, \bibinfo{person}{Junjie Chen}, \bibinfo{person}{Zhaoyang Yu}, {et~al\mbox{.}}} \bibinfo{year}{2021}\natexlab{}.
\newblock \showarticletitle{Identifying bad software changes via multimodal anomaly detection for online service systems}. In \bibinfo{booktitle}{{\em Proceedings of the 29th ACM Joint Meeting on the Foundations of Software Engineering(FSE)}}. \bibinfo{pages}{527--539}.
\newblock


\bibitem[\protect\citeauthoryear{Zhou, Peng, et~al\mbox{.}}{Zhou et~al\mbox{.}}{2018}]%
        {trainticket}
\bibfield{author}{\bibinfo{person}{Xiang Zhou}, \bibinfo{person}{Xin Peng}, {et~al\mbox{.}}} \bibinfo{year}{2018}\natexlab{}.
\newblock \showarticletitle{Fault analysis and debugging of microservice systems: Industrial survey, benchmark system, and empirical study}.
\newblock \bibinfo{journal}{{\em IEEE Transactions on Software Engineering\/}} \bibinfo{volume}{47}, \bibinfo{number}{2} (\bibinfo{year}{2018}), \bibinfo{pages}{243--260}.
\newblock


\bibitem[\protect\citeauthoryear{Zhou, Peng, et~al\mbox{.}}{Zhou et~al\mbox{.}}{2019}]%
        {MEPFL}
\bibfield{author}{\bibinfo{person}{Xiang Zhou}, \bibinfo{person}{Xin Peng}, {et~al\mbox{.}}} \bibinfo{year}{2019}\natexlab{}.
\newblock \showarticletitle{Latent error prediction and fault localization for microservice applications by learning from system trace logs}. In \bibinfo{booktitle}{{\em Proceedings of the 2019 27th ACM Joint Meeting on European Software Engineering Conference and Symposium on the Foundations of Software Engineering}}. \bibinfo{pages}{683--694}.
\newblock


\end{thebibliography}

\end{document}